\documentclass[superscriptaddress,pre,twocolumn,floatfix]{revtex4}

\usepackage{graphicx}
\usepackage{color}
\usepackage{rotating}
\usepackage{amsmath}
\usepackage{amsfonts}
\usepackage{amssymb}
\usepackage{enumerate}
\usepackage{longtable}
\usepackage{dcolumn}
\usepackage{bm}
\usepackage{epsfig}
\usepackage{hyperref}
\begin{document}
\title{Spatiotemporal correlations of handset-based service usages}
\author{Hang-Hyun Jo}
\email{hang-hyun.jo@aalto.fi}
\affiliation{Department of Biomedical Engineering and Computational Science, Aalto University School of Science, P.O. Box 12200, Finland}
\author{M\'arton Karsai}
\affiliation{Department of Biomedical Engineering and Computational Science, Aalto University School of Science, P.O. Box 12200, Finland}
\author{Juuso Karikoski}
\affiliation{Department of Communications and Networking, Aalto University School of Electrical Engineering, P.O. Box 13000, Finland}
\author{Kimmo Kaski}
\affiliation{Department of Biomedical Engineering and Computational Science, Aalto University School of Science, P.O. Box 12200, Finland}

\date{\today}

\begin{abstract}
We study spatiotemporal correlations and temporal diversities of handset-based service usages by analyzing a dataset that includes detailed information about locations and service usages of 124 users over 16 months. By constructing the spatiotemporal trajectories of the users we detect several meaningful places or contexts for each one of them and show how the context affects the service usage patterns. We find that temporal patterns of service usages are bound to the typical weekly cycles of humans, yet they show maximal activities at different times. We first discuss their temporal correlations and then investigate the time-ordering behavior of communication services like calls being followed by the non-communication services like applications. We also find that the behavioral overlap network based on the clustering of temporal patterns is comparable to the communication network of users. Our approach provides a useful framework for handset-based data analysis and helps us to understand the complexities of information and communications technology enabled human behavior.
\end{abstract}


\maketitle

\section{Introduction}
\label{sec:intro}

Understanding macroscopic socio-economic phenomena of a large number of individuals has been extensively studied by means of social, physical, and computational sciences~\cite{Goyal2009,Castellano2009,Lazer2009}. Recent access to large-scale digital datasets on human dynamics and social interaction has enabled us to quantitatively investigate the structure and dynamics of human communication networks. Indeed, researchers have studied various datasets, ranging from email and mobile phone communications to social network services, e.g. Twitter and Facebook~\cite{Eckmann2004,Barabasi2005,Onnela2007a,Kwak2010,Lewis2008,Kovanen2011,Jo2012,Karsai2012}. Mobile phones or handsets are now actively utilized to accurately measure or sense human behavior because the handsets equipped with a variety of sensors, including GPS and WiFi, are carried around by the users everyday and all day through. Highly resolved location data collected from handsets have been recently used to uncover human mobility patterns~\cite{Gonzalez2008,Candia2008,Wang2009,Song2010a,Song2010b,Eagle2006,Eagle2009,Krings2009,Bagrow2012}. The reliability of data collected from handsets, i.e. ``behavioral'' data, was tested in the serial studies conducted within the frame of MIT's Reality Mining project~\cite{Eagle2006,Eagle2009,Aharony2011}. It was shown that the behavioral data are at least comparable to self-report survey data in terms of friendship network and even capturing information that self-reports are missing~\cite{Eagle2009}.

The handset usage patterns are known to be diverse among users when measured by the number or duration of the phone sessions and by the amount of data received, to name a few~\cite{Falaki2010,Soikkeli2011b}. Within the individual handset usage patterns, temporal inhomogeneities due to circadian and weekly cycles were also reported~\cite{Jo2012}, which are in close relation to the spatial inhomogeneities, such as nighttime at home and daytime in office. Therefore, for conducting a comprehensive study, it is important to identify the context characterizing the situation of handset user, and then to understand how the context affects service usage patterns~\cite{Dey2001,Verkasalo2009,Soikkeli2011a,Soikkeli2011b,Karikoski2011a}. However, it is only very recently when the effect of context on the handset-based service usages was investigated. But so far the analysis has been conducted mostly at the aggregate level, while the temporal diversities of service usage among users have been ignored~\cite{Karikoski2011a}. 

In this paper, we study spatiotemporal correlations of the service usage patterns of individual users by analyzing a handset-based dataset. This dataset was collected from 124 users' handsets for over 16 months as a part of the OtaSizzle project at Aalto University, Finland~\cite{sizl}. A software installed on handsets collected information about the handset's locations and usages of various services, including web domain visits, applications, emails, voice calls, and short message services, with the resolution of seconds in time and mobile network base stations spatially. After constructing spatiotemporal trajectories of the users we identify several contexts that are meaningful to them by using the context detection method~\cite{Soikkeli2011a}. Other methods include, for example, places of interest or meaningful locations~\cite{Montoliu2010,Nurmi2006} and eigenmode analysis~\cite{Eagle2009b,Reades2009,Park2010}. Then, we find correlations between the spatiotemporal trajectories and the service usage patterns. We observe the similarity and diversity in temporal patterns of the service usages and discuss their temporal correlations, time-ordering behavior between services, and behavioral overlap network based on the clustering results. Our approach provides a useful framework for handset-based data analysis, and hence it would be important for better design of information and communications technology (ICT) enabled social environments and services. 

This paper is organized as follows. In Section~\ref{sec:data} we describe the data collection and preparation methods. In Section~\ref{sec:detect} several contexts for each user are identified by means of the context detection method applied to user's spatiotemporal trajectory. In Section~\ref{sec:correl} we uncover the spatiotemporal correlations and the similarity and diversity in temporal patterns of the service usages. Finally, we summarize the results with concluding remarks in Section~\ref{sec:summary}.

\section{Handset-based dataset}
\label{sec:data}

\subsection{Data collection method}

The handset-based dataset in this study was collected by the MobiTrack software installed on Nokia Symbian smartphones of 183 participants or users from September 2009 to December 2010, i.e. for a period spanning about 16 months. All users were students and staff members of Aalto University, Finland and identified as early adopters of mobile phones and services~\cite{Karikoski2012}. The dataset was anonymized so that no personal information of the users could be obtained. We consider only 124 users with the overall duration of handset usage longer than 30 days, see Section~\ref{sec:detect} for details. 

\begin{figure}[!t]
\includegraphics[width=\columnwidth]{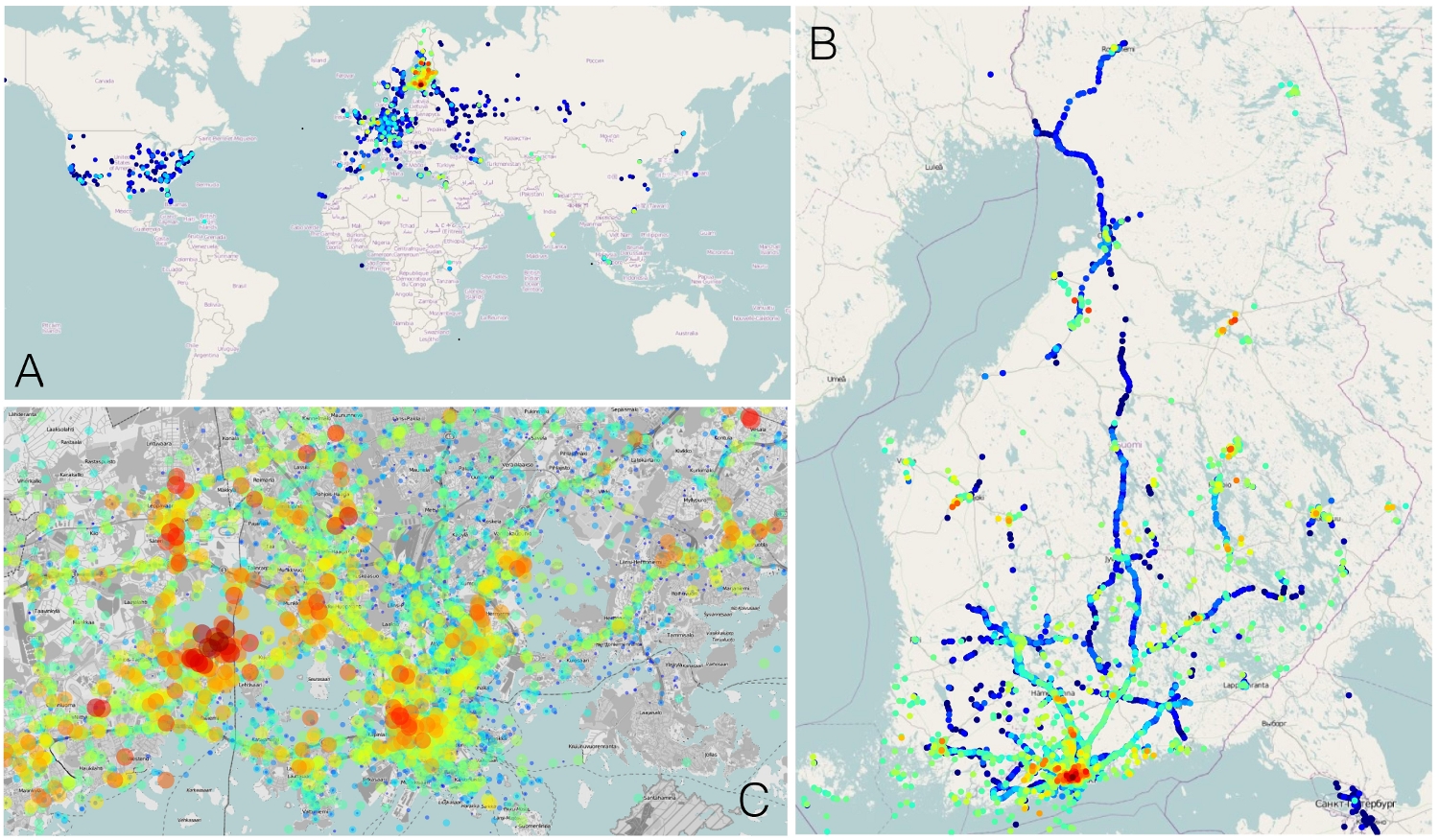}
\caption{Recording frequencies at mobile network base stations by all users. (a) Over the world, (b) in Finland, and (c) in the Helsinki municipal area. The higher frequency is denoted by the warmer color. In (c) the size of circle is logarithmic in frequency.}
\label{fig:maps}
\end{figure}

The dataset consists of two kinds of information: locations and service usages. The resolution of locations is limited to the physical area covered by each mobile network base station, i.e. cell, denoted by $c$. Whenever the handset is connected to a new cell or otherwise every half an hour, the identifier of the cell connected by the handset was recorded with a timestamp $t$ with one second resolution. Each cell can be located in the geographical space with a unique pair of latitude and longitude. The geographic information for cells and the maps used in Figs.~\ref{fig:maps} and~\ref{fig:contexts} were collected as a part of the OpenNetMap project and from open databases~\cite{opennetmap,opencellid,locationapi}. For all users we have 5596041 records at 99206 different cells. Although only $29.0\%$ of cells could be located in the geographical space, they correspond to $91.3\%$ of records. Figure~\ref{fig:maps} shows all located cells over the world, in Finland, and in the Helsinki municipal area. In this way, the detailed spatiotemporal trajectory of each user could be constructed in terms of a sequence of cell records $\{(c_k,t_k)\}$, where $k$ denotes the ordered index of record.

For service usage data we consider five services: web domain visit (web), application (app), email, voice call (call), and short message service (SMS). Each service usage or event was recorded with a timestamp with one second resolution together with service-specific relevant information. In the case of web domain visits, a URL (Uniform Resource Locator) was extracted and recorded whether it was visited via browser or widget. Only the applications visible in the foreground of the handset were recorded so that no process or application running in the background was considered. The records of communication services, such as email, call, and SMS, include the information on whether the user was an initiator or receiver of the communication event, and on the communication partner if available. For more information regarding the data collection method, see~\cite{Karikoski2012}.

\begin{figure}[!t]
\includegraphics[width=.9\columnwidth]{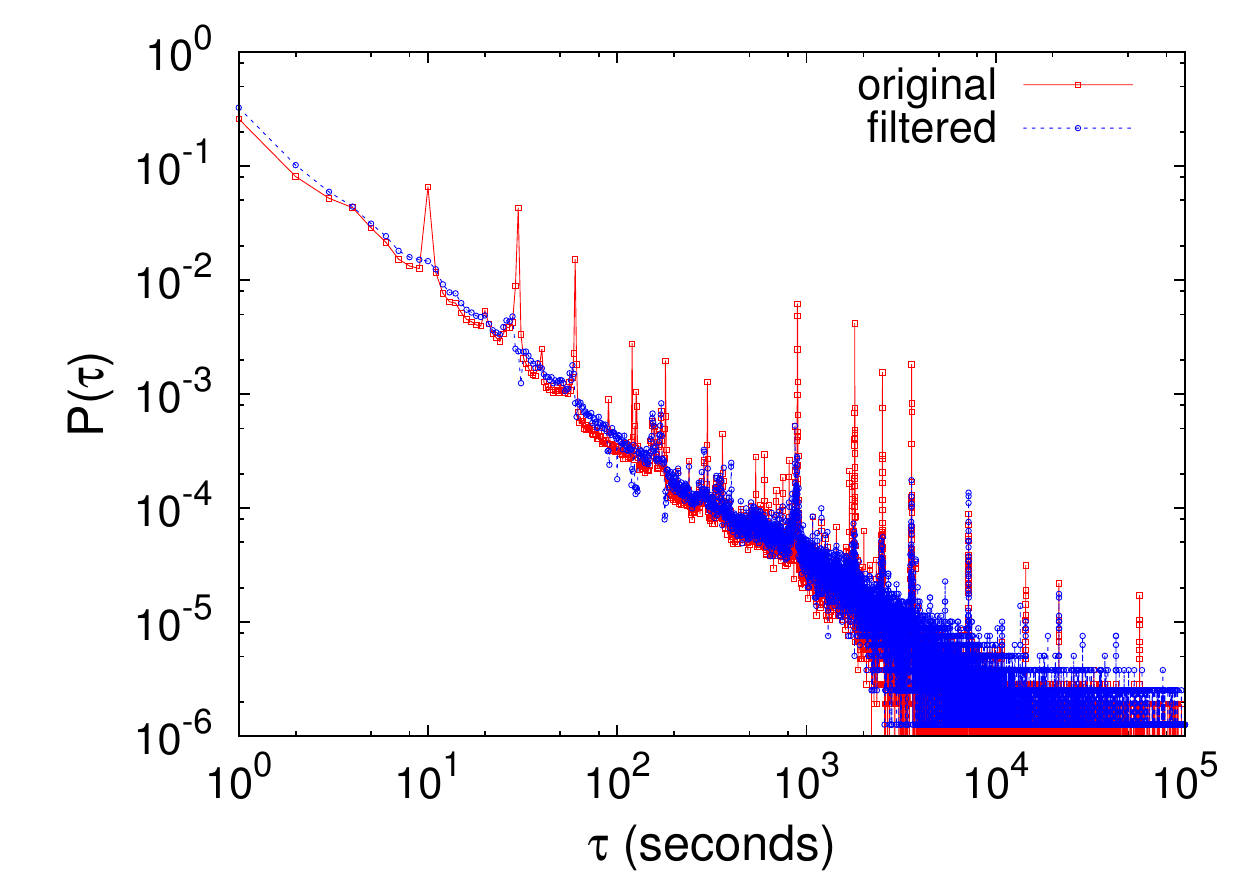}
\caption{Original and filtered distributions of inter-event time for web domain visits by all users. The inter-event time is defined as the time interval between consecutive web domain visits by the same user. The peaks due to automatic events by the browser have been successfully suppressed after filtering.}
\label{fig:distrInterWeb}
\end{figure}

\subsection{Data preparation method}

The service usage dataset contains events mostly generated by users but it also contains automatic events by the operating system of the handsets. In order to observe the pure human behavior, we systematically filtered out these automatic events. However, some spurious regularities still remain in the web dataset. In the cases of google.com, facebook.com and so on, once a web is connected, the browser might visit the same web automatically for periodic updates and synchronization of accounts until the web is disconnected. To resolve this issue, we obtain the distribution of inter-event time $\tau$, defined as the time interval between consecutive web domain visits by the same user. Several sharp peaks at specific inter-event times are found, where each peak is mostly related to the single webpage. We remove all the events leading to those inter-event times, except for the event trains consisting of only two events with $\tau=10$ seconds. It is because some trains with only two events separated by $10$ seconds can also be generated by users. As new regularities become visible after filtering, we apply this method recursively until the peaks are suppressed considerably, leading to an approximately $25\%$ of entire events removed. Figure~\ref{fig:distrInterWeb} shows that this filtering method for web dataset does not change the overall characteristics of the inter-event time distribution.

We also ignore some user-generated application events associated with other service usages, corresponding to $17\%$ of entire events. For example, the user opens the messaging application when sending or receiving SMSs. These application events might lead to artificial correlations between different service usages. In addition, corrupted events, less than $0.1\%$ of the whole dataset, have been ignored or manually corrected. Finally, we have 792971 web domain visits, 433726 application events, 17976 emails, 79779 calls, and 79283 SMSs in the service usage dataset. 

\begin{figure}[!t]
\includegraphics[width=.9\columnwidth]{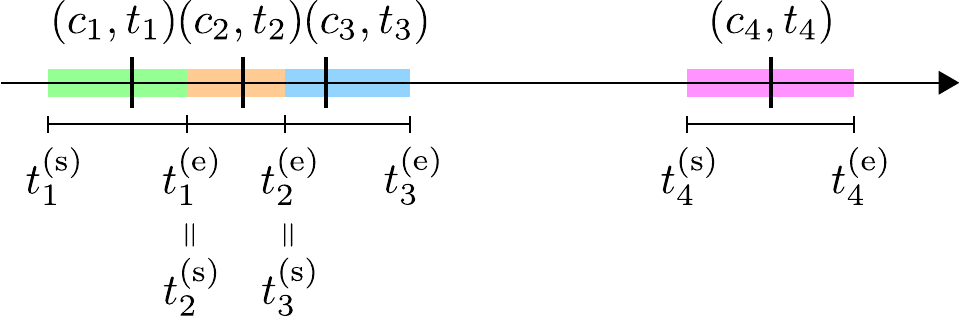}
\caption{ 
Schematic diagram for deriving temporal boundaries $\{(c_k,t^{\rm (s)}_k,t^{\rm (e)}_k)\}$ (colored boxes) from a sequence of cell records $\{(c_k,t_k)\}$ (vertical black lines). It is assumed that the user stays in the cell $c_k$ from the moment of $t^{\rm (s)}_k$ to $t^{\rm (e)}_k$. See the text for details.}
\label{fig:timeboundary}
\end{figure}

\section{Context detection from spatiotemporal pattern}
\label{sec:detect}

In order to detect the contexts for each user, we construct the user's spatiotemporal trajectory from a sequence of cell records $\{(c_k,t_k)\}$. It is necessary to infer the user's location between consecutive timestamps of cell records. From a sequence of cell records, we derive the temporal boundaries $\{(c_k,t^{\rm (s)}_k,t^{\rm (e)}_k)\}$ for the user's trajectory, implying that the user stays within the area covered by cell $c_k$ from the moment of $t^{\rm (s)}_k$ to $t^{\rm (e)}_k$, see Fig.~\ref{fig:timeboundary}. It is assumed that the user stays in the cell $c_k$ till $t^{\rm (e)}_k=\tfrac{1}{2}(t_k+t_{k+1})$ and then in the cell $c_{k+1}$ from $t^{\rm (s)}_{k+1}=t^{\rm (e)}_k$ when $t_{k+1}-t_k\leq 2t_c$. Here we set $t_c$ as half an hour, i.e. the time interval for regular cell recording. The time interval between consecutive timestamps longer than $2t_c$ implies that the handset may be turned off, used in offline or airplane mode, or not able to detect any cell nearby. If $t_{k+1}-t_k>2t_c$, the user is considered to stay in the cell $c_k$ till $t^{\rm (e)}_k=t_k+t_c$ and in the cell $c_{k+1}$ from $t^{\rm (s)}_{k+1}=t_{k+1}-t_c$. Hence, the location is unknown between $t^{\rm (e)}_k$ and $t^{\rm (s)}_{k+1}$. Then, the total time spent, i.e. duration, in each cell $c$ is obtained as follows:
\begin{equation}
    d_c=\sum_{\{k| c_k=c\}} (t^{\rm (e)}_k-t^{\rm (s)}_k).
\end{equation}
If the sum of durations in all the recorded cells, $D\equiv\sum_c d_c$, is less than 30 days, that user is not considered for the further analysis, leading to 124 available users. The average and standard deviation of $D$ for available users are $121\pm 63$ days.

\begin{figure}[!t]
\includegraphics[width=\columnwidth]{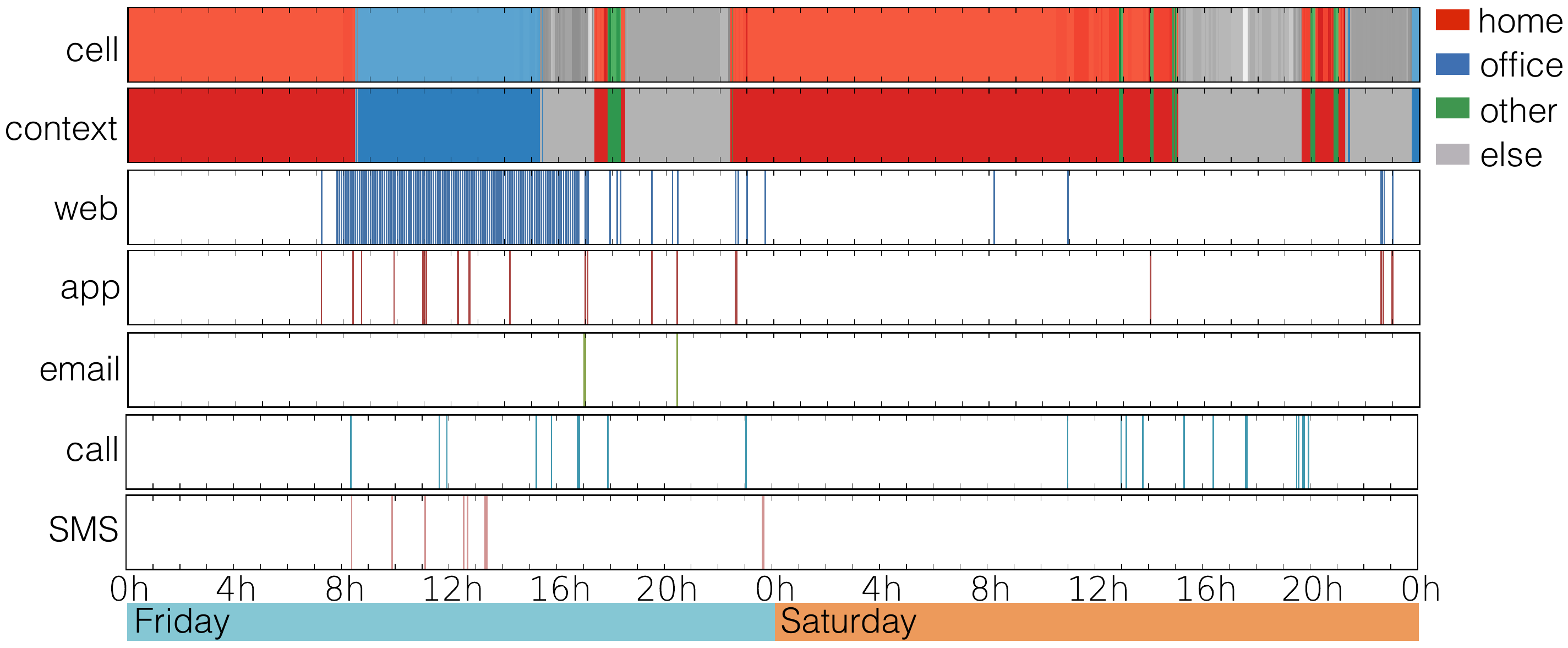}
\caption{Locations and service usage patterns of a sample user 81 during typical Friday and Saturday. The first and second rows represent cells and contexts assigned to cells. Home, Office, Other meaningful place, and Elsewhere are denoted by red, blue, green, and gray colors, respectively. Different depths of the same color indicate the different cells belonging to the same context. Service usage events are denoted by vertical lines in the rows of web, app, email, call, and SMS (from the third to the bottom).}
\label{fig:user81}
\end{figure}

In addition, we observe back and forth changes in a short time span between two cells covering the neighboring areas. It can occur even without any real movement of the handset if the handset is located at the boundary of two neighboring cells. To filter out this noisy behavior, the involved cells can be clustered by a sandwich clustering method~\cite{Soikkeli2011a}. Here we consider only one type of sandwich with four records involving two cells, i.e. $c_k=c_{k+2}\neq c_{k+1}=c_{k+3}$ with $t^{\rm (e)}_l-t^{\rm (s)}_l\leq t_c$ for $l=k,\cdots,k+3$. Whenever this type of sandwich is detected, every $c_k$ in the temporal boundaries is replaced by or merged into $c_{k+1}$ if $d_{c_{k+1}} >d_{c_k}$, and vice versa. Consequently, some geographically neighboring cells can be clustered into one representative cell, which from now on will be considered equally with normal cells. For example, the first row in Fig.~\ref{fig:user81} shows the user 81's temporal boundaries during typical Friday and Saturday. Note that clustering cells for one user is independent of other users' records.

\begin{figure}[!t]
\includegraphics[width=.9\columnwidth]{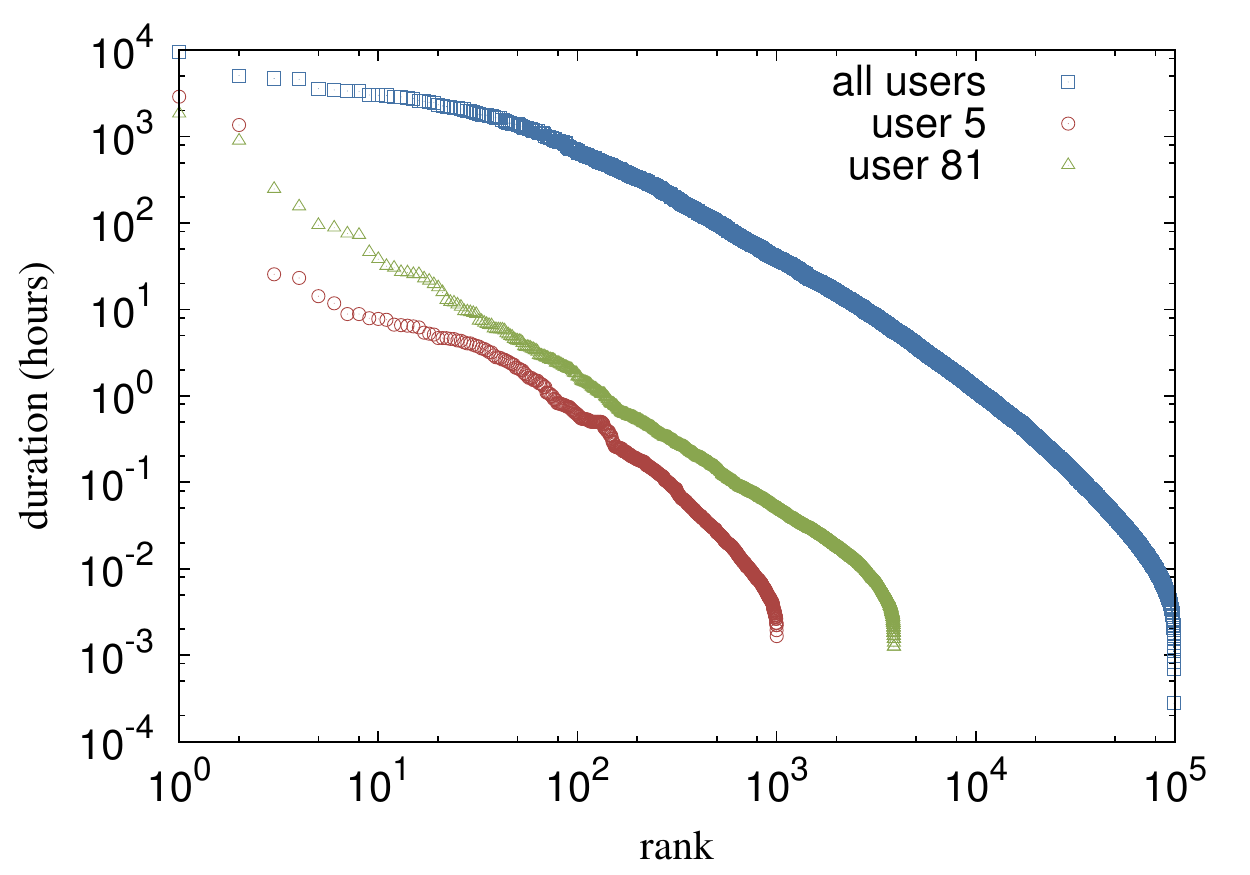}
\caption{ 
Rank curves of cells for all users and for sample users 5 and 81. Here the rank curve is defined as an ordered sequence of durations in cells.}
\label{fig:durationRank}
\end{figure}

\begin{figure}[!t]
\includegraphics[width=\columnwidth]{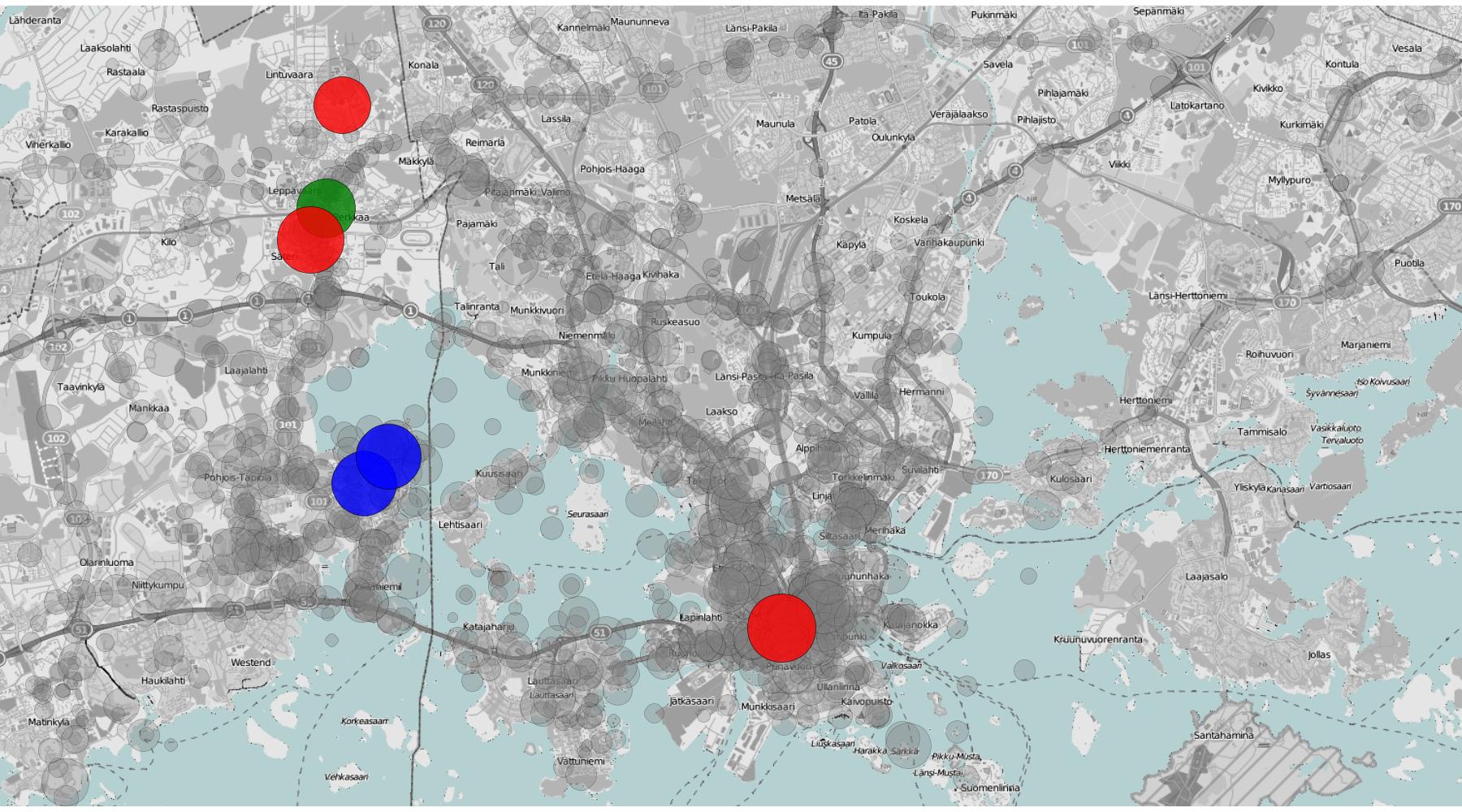}
\caption{Contexts detected for sample users 5 and 81 in a map of Helsinki municipal area. Each cell is represented by the circle with radius according to the duration in that cell. The cells are identified as either Home (red), Office (blue), Other meaningful place (green), or Elsewhere (gray).}
\label{fig:contexts}
\end{figure}

We find spatiotemporal inhomogeneities of the trajectories of handsets on the individual basis as well as at the aggregate level. As an illustrative example, we obtain the rank curve $d(r)$, defined as the duration in the $r$th cell $c$ in a descending order according to $d_c$. The rank curve for all users is highly skewed, such that the first few cells, including one in Otaniemi campus of Aalto University, were visited for more than a few months while $88.9\%$ of cells were visited for less than one hour, as shown in Fig.~\ref{fig:durationRank}. The same inhomogeneities are also observed for individual users. For example, the rank curves for users 5 and 81 are shown in Fig.~\ref{fig:durationRank}, who were selected to show the representative behavior.

The heavily visited cells are supposed to cover meaningful places to the handset user, such as home and office. Since the service usage patterns might be affected by the different characteristics of meaningful places, it is important to identify the context characterizing the situation of user. Here the context is preferred to the meaningful place because the time and place of handset usage are not independent but correlated, e.g. nighttime at home and daytime in office~\cite{Soikkeli2011a}. Each cell will be detected as one of five contexts, such as Home, Office, Other meaningful place (Other), Elsewhere (Else), and Abroad. One context can be assigned to several cells. The identifier of a cell contains the mobile country code (MCC), by which Abroad context is assigned to the cells out of Finland. For the cells within Finland, we obtain more detailed durations for each cell $c$: 
\begin{enumerate}
    \item duration on weekdays ($d_{c,{\rm wd}}$),
    \item duration on weekdays between 0 AM and 6 AM ($d_{c,0-6}$), and
    \item duration on weekdays between 10 AM and 4 PM ($d_{c,10-16}$).
\end{enumerate}

Now we describe criteria for assigning contexts except for Abroad. A cell is detected as Elsewhere (Else) if the duration in that cell is negligible to the total duration as
\begin{equation}
    d_c/D<t_{\rm elsewhere}=0.02.
\end{equation}
For example, Else is assigned to the cells along the highways. The threshold value of $t_{\rm elsewhere}$ has been determined in order to leave only $0.2\%$ of cells, i.e. $3.73$ cells per user, for other contexts. A cell is detected as Office if the user spends a considerable time in that cell during the working time on weekdays as
\begin{equation}
    d_{c,{\rm wd}}/d_c>t_{\rm weekday}=0.8
\end{equation}
and 
\begin{equation}
    d_{c,10-16}/d_{c,{\rm wd}}>t_{\rm worktime}=0.5.
\end{equation}
With above threshold values, at least one Office has been detected for more than half of the users. Note that most users were students so that they might not have any regular places to visit during the working time. Next, Home is assigned to a cell if the user spends a considerable time in that cell for nighttime and free time, i.e. the remaining time except for the working time, on weekdays as
\begin{equation}
    d_{c,0-6}/d_{c,{\rm wd}}>t_{\rm nighttime}=0.1
\end{equation}
and 
\begin{equation}
    d_{c,10-16}/d_{c,{\rm wd}}<t_{\rm freetime}=0.3.
\end{equation}
With above threshold values, at least one Home has been detected for all users except for two of them. Many users turn out to have more than one Home, such as user's own home and his/her parent's home. Finally, the remaining cells are detected as Other meaningful place (Other). Figure~\ref{fig:contexts} shows the locations of detected contexts for sample users in the Helsinki municipal area. We put two sample users' contexts together to avoid privacy issues. 

\begin{figure}[!t]
\includegraphics[width=.9\columnwidth]{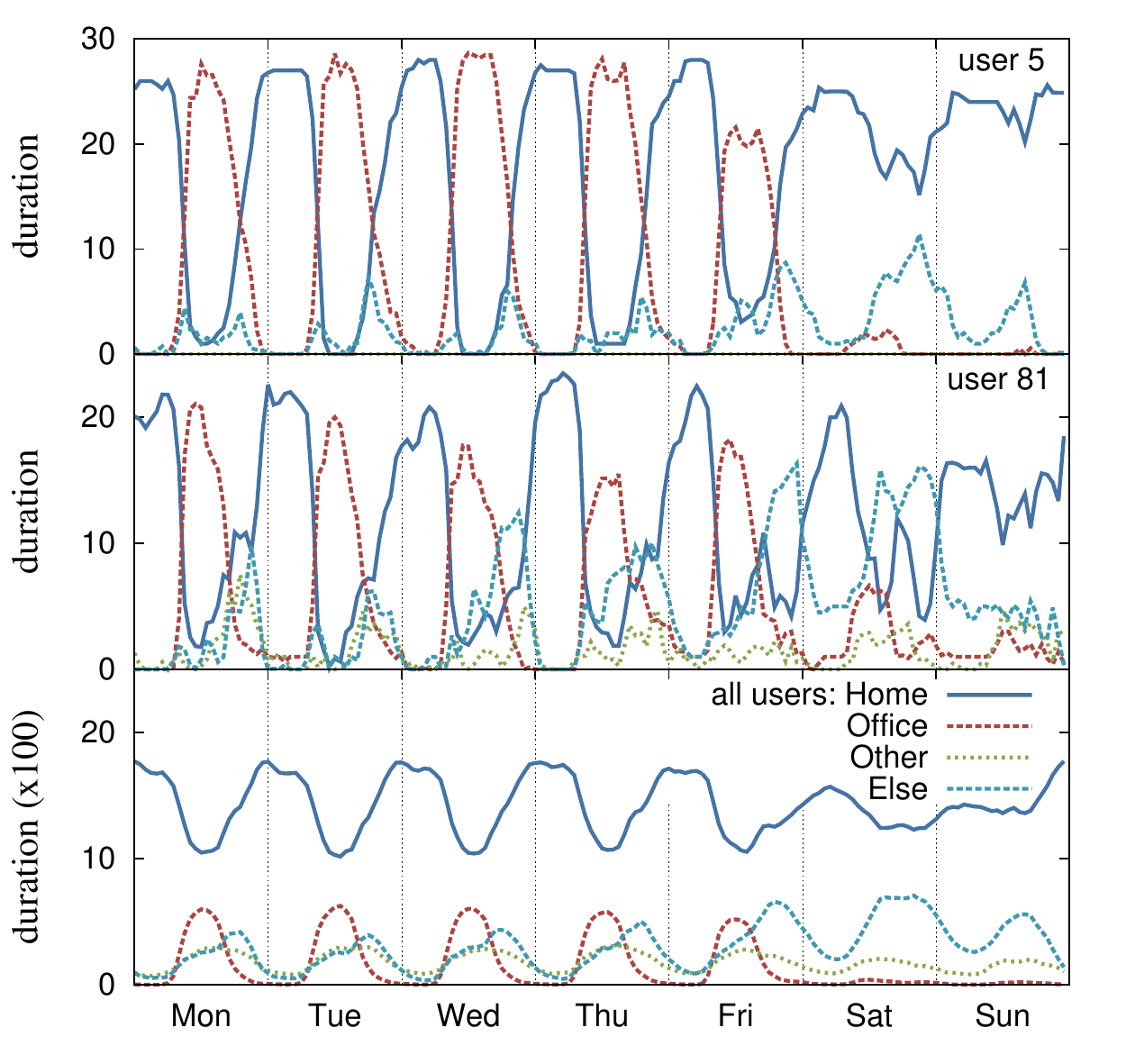}
\caption{ 
Weekly patterns of duration in hours for the different contexts for users 5 and 81 and for all users (from top to bottom). The typical weekly cycles of humans are observed.}
\label{fig:contextDurationWeek}
\end{figure}

Our context detection method is validated by weekly patterns of duration for different contexts obtained for sample users and at the aggregate level, as depicted in Fig.~\ref{fig:contextDurationWeek}. For example, the user 5 without Other detected shows a very regular pattern, especially on weekdays, i.e. at Home in nighttime, in Office during the working time, and at Else when moving between Home and Office. Weekly patterns of user 81 are comparable to the temporal boundaries in terms of detected contexts, as depicted in the second row in Fig.~\ref{fig:user81}. Weekly patterns of duration aggregated over all users show the overall behavior. Durations at Home, Office, Other, and Else account for $66.8\%$, $7.0\%$, $8.5\%$, and $14.0\%$ of the total duration of all users, respectively.

\section{Spatiotemporal correlations of service usages}
\label{sec:correl}

We investigate correlations between users' spatiotemporal trajectories and their service usage patterns. Here five services, such as web domain visit (web), application (app), email, voice call (call), and short message service (SMS), are considered and each service is denoted by $s$. The spatiotemporal correlation of service usages for user $i$ is fully characterized by the number of events corresponding to the service $s$ in the cell $c$ and at time $t$, denoted by $n_{is}(c,t)$. For gaining contextual understanding of correlations we consider the contexts instead of cells, i.e. $n_{is}(C,t)=\sum_c n_{is}(c,t)$, where the summation is over $c$ detected as context $C$. 

\begin{figure*}[!t]
\includegraphics[width=1.9\columnwidth]{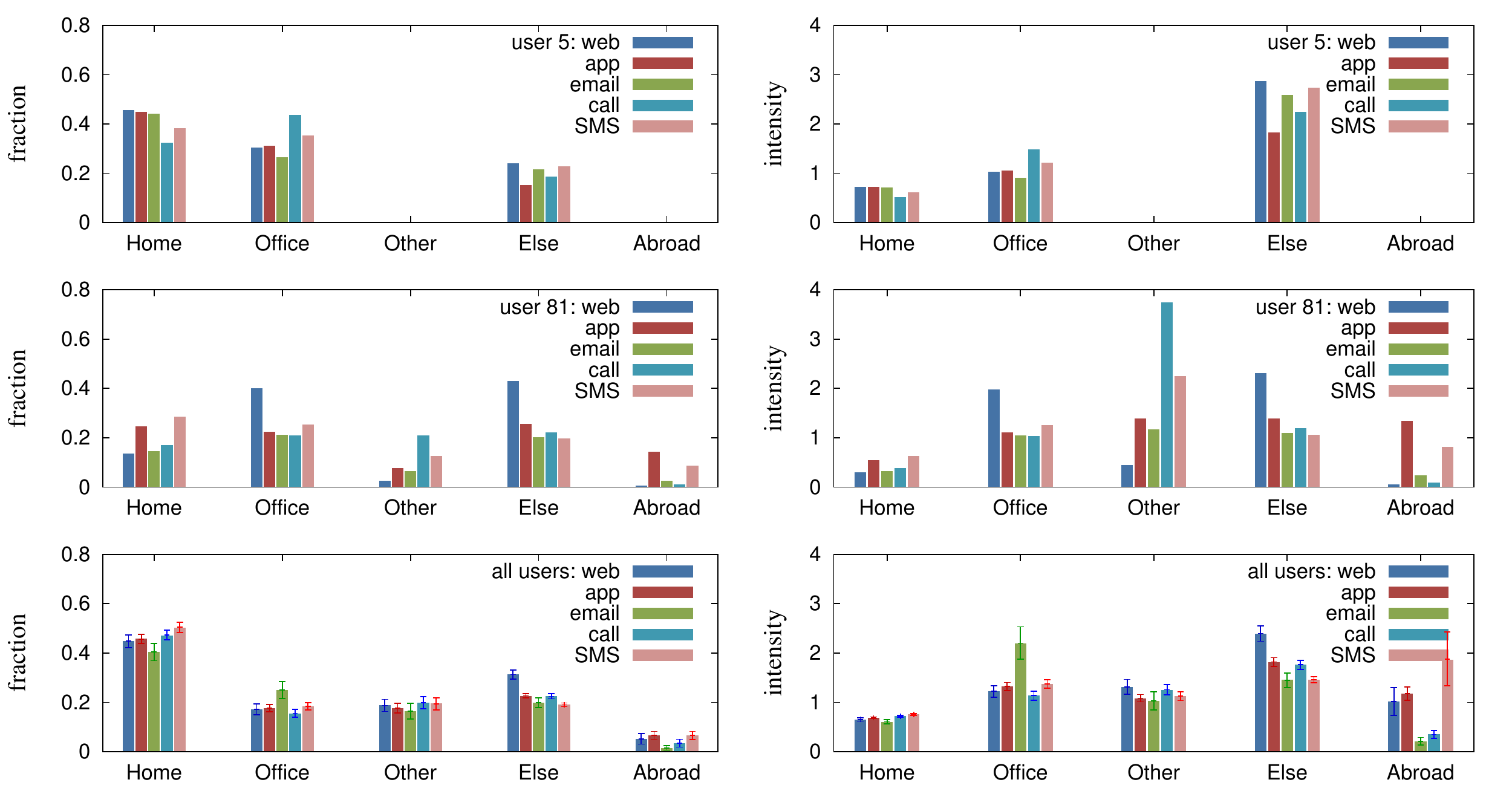}
\caption{ 
Contextual correlations of service usages for users 5 and 81 and for all users (from top to bottom). Fractions (left) and intensities (right) of service usages are defined in Eqs.~(\ref{eq:usageFraction}) and (\ref{eq:usageIntensity}), respectively. Standard errors are also provided for the user-averaged statistics.}
\label{fig:actCellContext}
\end{figure*}

\subsection{Contextual correlations of service usages}

We first focus on the contextual correlations of service usages with $n_{is}(C)=\sum_t n_{is}(C,t)$. Since services have qualitatively different characteristics, the numbers of events of different services cannot be directly compared to each other but only in terms of fractions and intensities of usages. The fraction of service usage is defined as follows
\begin{equation}
    \label{eq:usageFraction}
    f_{is}(C)=\frac{n_{is}(C)}{\sum_C n_{is}(C)}.
\end{equation}
Figure~\ref{fig:actCellContext}~(left) shows the fractions for sample users 5 and 81 as well as their means over all users with standard errors, measured by the bootstrap method. The handset of user 5 has never been abroad and no Other context is detected. For this user all service usages are more active at Home and Office than at Else, which is very different from the service usage patterns of user 81. Due to the diversity of the service usage patterns among users, any general conclusion cannot be made on the individual basis. However, by looking at the means with standard errors, it is found that all service usages are the most active at Home, while they are relatively inactive for other contexts. Given the aggregate durations for different contexts obtained in the Section~\ref{sec:detect}, this finding can be explained such that the longer duration for some context means the higher chance for service usage.

Accordingly, instead of the fractions of service usages we consider those divided by the corresponding durations as follows:
\begin{equation}
    \label{eq:usageIntensity}
    I_{is}(C)=\frac{n_{is}(C)}{\sum_C n_{is}(C)}\cdot\frac{\sum_C d_{iC}}{d_{iC}},
\end{equation}
where $d_{iC}$ denotes the duration of user $i$ for context $C$. The results are shown in Fig.~\ref{fig:actCellContext}~(right). Despite of the diversity among users, the means of intensities of different services for the same context have to some extent similar values. The large mean of intensity of email usage in Office might be due to the fact that users prefer emails to calls or SMSs in classes or laboratories during the working time. The large mean of intensity of web usage at Else could be the result of users killing time by surfing the webpages while on the move. One could also say that users while abroad tend to use SMSs more than other communication services. Finally, for all services, only the means of intensity at Home turn out to be less than 1 and most inactive, which could be partly because users have many other activities to do at Home.

\subsection{Temporal correlations and time-ordering of service usages}

We turn to analyze the temporal correlations of service usages in terms of $n_{is}(t)=\sum_C n_{is}(C,t)$, where the summation is over all contexts with one exception, Abroad. It is because the service usage abroad cannot be considered as normal, as shown in Fig.~\ref{fig:actCellContext}. We first obtain weekday and weekend patterns of service usages as 
\begin{eqnarray}
n^{\rm wd}_{is}(t)&=&\sum_{k} n_{is}(t+kT_d),\\
n^{\rm we}_{is}(t)&=&\sum_{k'} n_{is}(t+k'T_d)
\end{eqnarray}
for $0\leq t<T_d$ with $T_d=1$ day. Here $k$ and $k'$ denote the indexes of weekdays and weekends, respectively. The weekday and weekend event rates of service $s$ for user $i$ are defined as
\begin{eqnarray}
\label{eq:shortPatternWeekday}
\rho^{\rm wd}_{is}(t)&=&\frac{a n^{\rm wd}_{is}(t)}{\sum_t [a n^{\rm wd}_{is}(t)+a' n^{\rm we}_{is}(t)]},\\
\label{eq:shortPatternWeekend}
\rho^{\rm we}_{is}(t)&=&\frac{a' n^{\rm we}_{is}(t)}{\sum_t [a n^{\rm wd}_{is}(t)+a' n^{\rm we}_{is}(t)]},
\end{eqnarray}
where $a=1/5$ and $a'=1/2$ are weights for normalization. In addition we obtain the weekday and weekend event rates averaged over all users.

\begin{figure}[!t]
\includegraphics[width=.9\columnwidth]{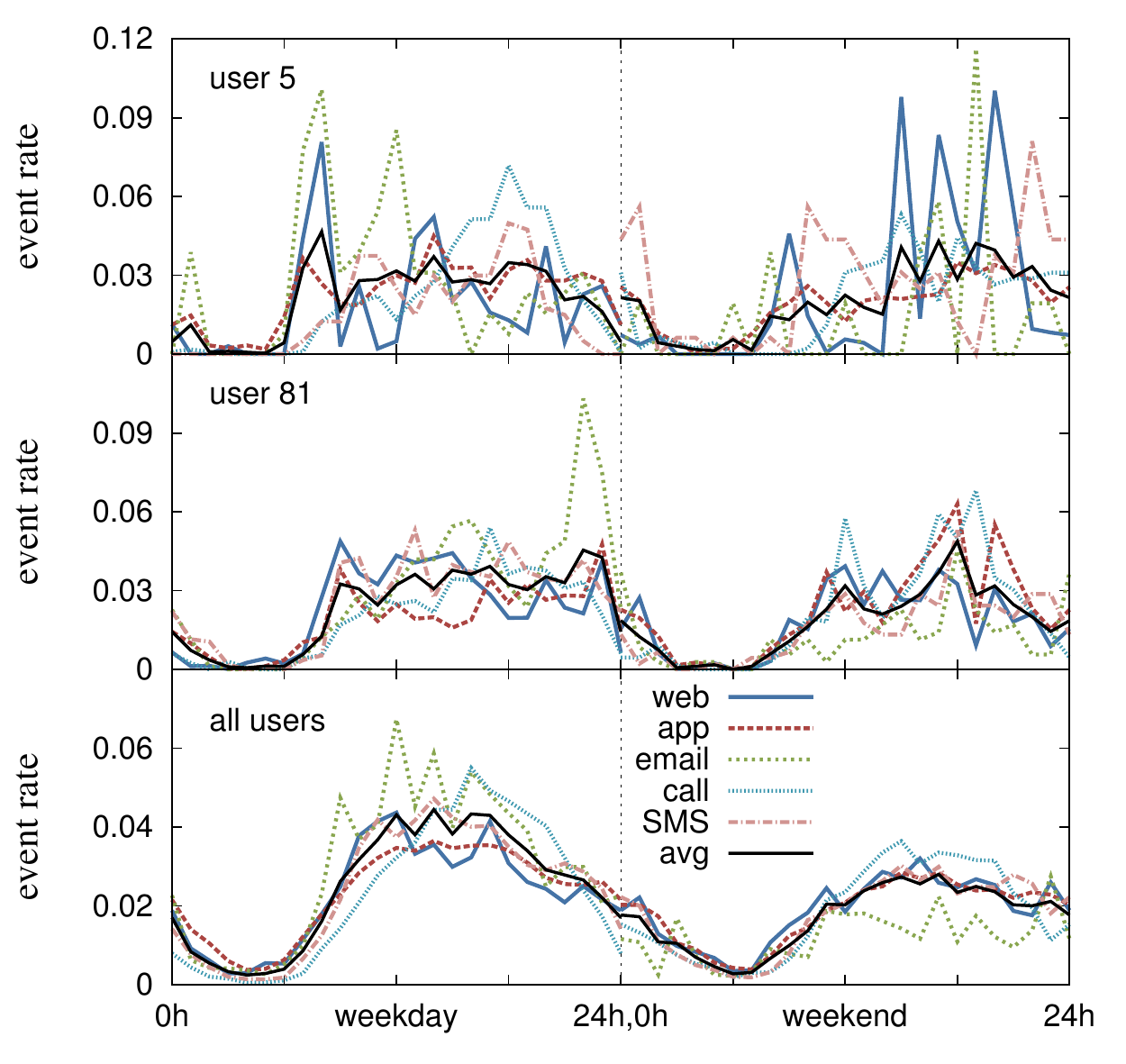}
\caption{
Weekday and weekend service usage patterns for users 5 and 81 and for all users (from top to bottom), showing the similarity and diversity among users. The bin size was set to one hour.}
\label{fig:actShortPattern}
\end{figure}

\begin{figure}[!h]
\includegraphics[width=.9\columnwidth]{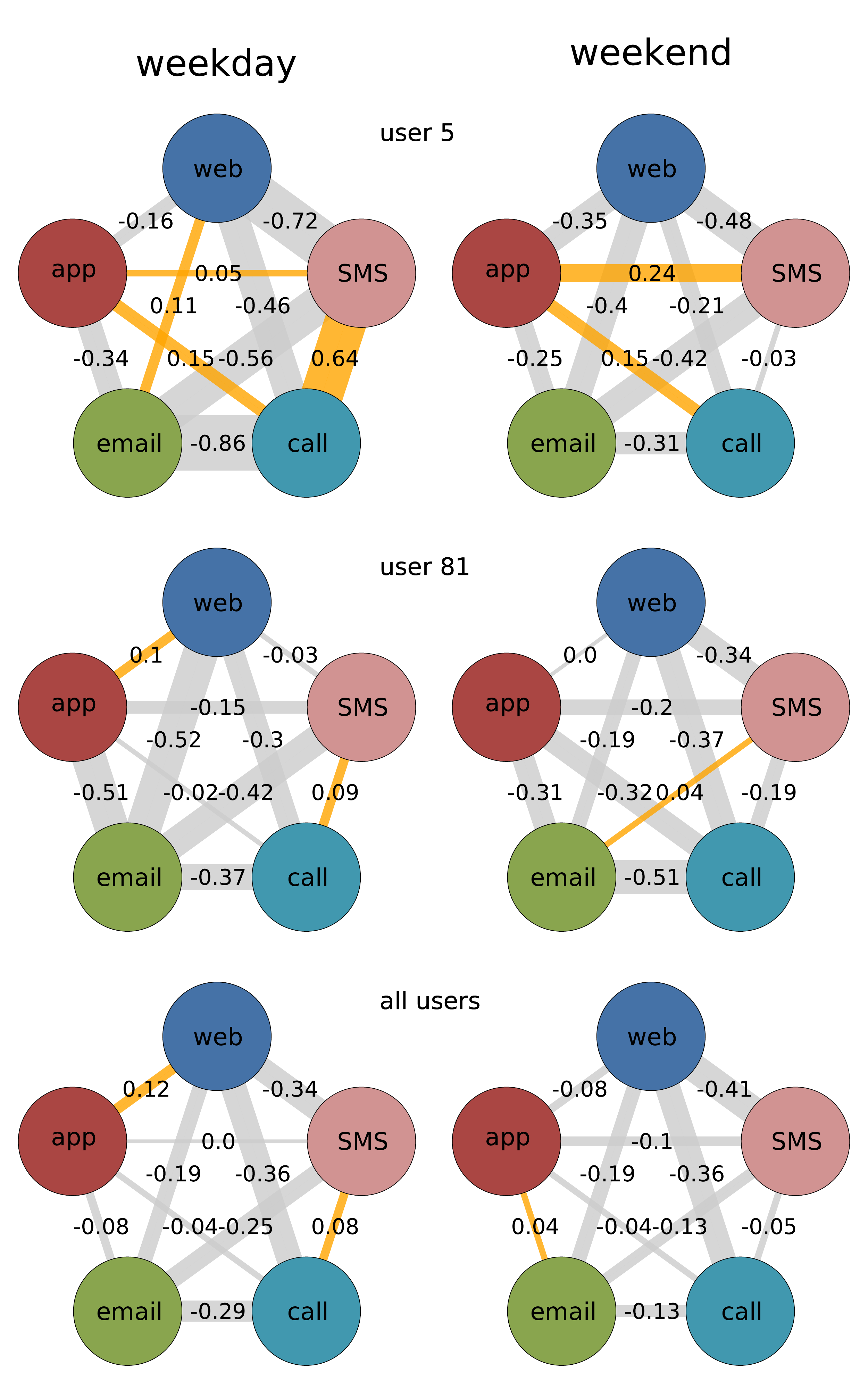}
\caption{
Pearson correlation coefficients among service usages for users 5 and 81 and for all users (from top to bottom), obtained from weekday (left) and weekend (right) event rates. Positive and negative correlations are represented by orange and gray lines, respectively.}
\label{fig:actWeekCorrel}
\end{figure}

\begin{figure}[!h]
\includegraphics[width=\columnwidth]{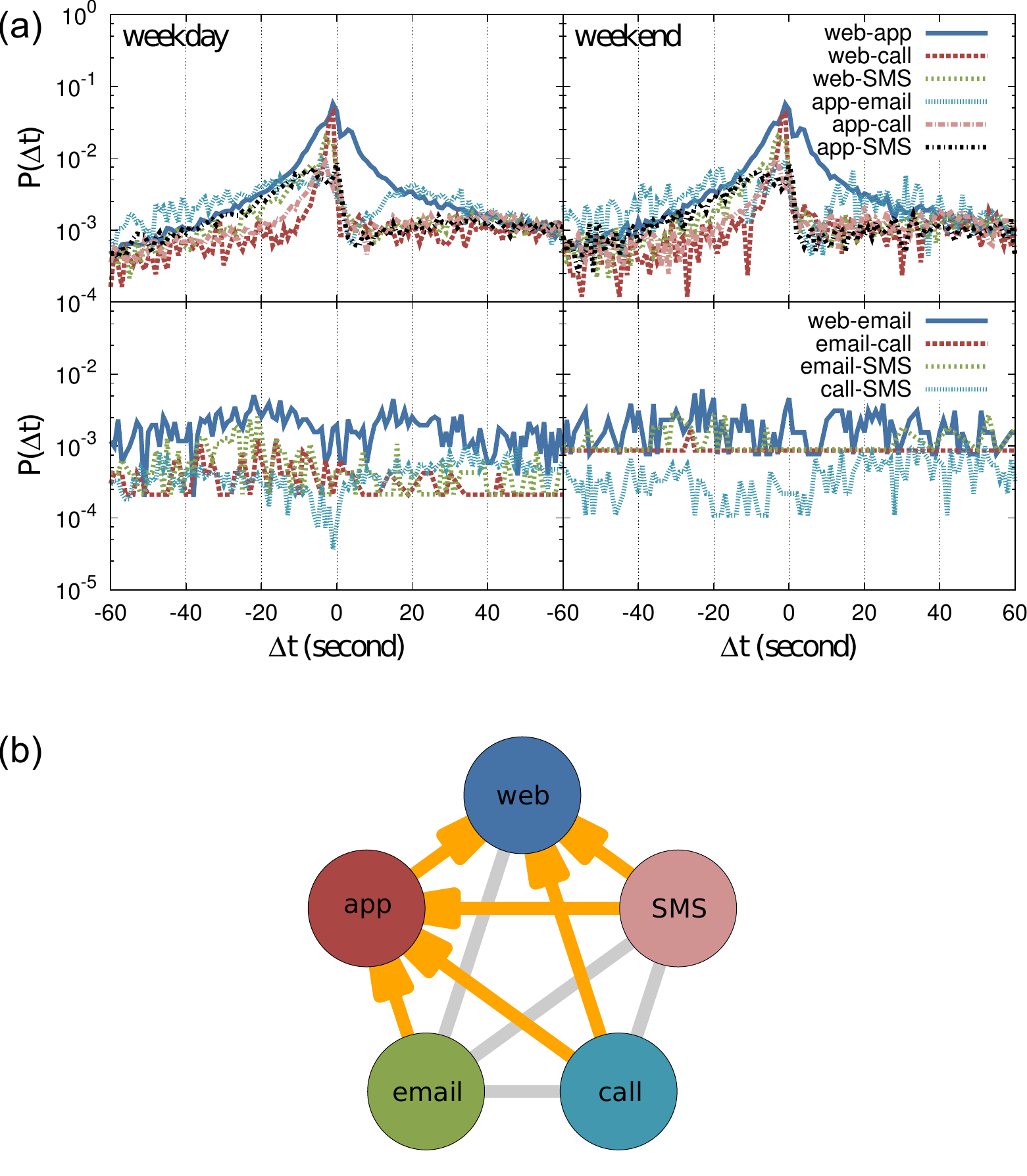}
\caption{(a) Distributions of time interval $\Delta t_{ss'}$ between consecutive events of different services $s$ and $s'$. (b) Diagram for time-ordering behavior between services based on the distributions of time interval.}
\label{fig:crossCorrel}
\end{figure}

In Fig.~\ref{fig:actShortPattern} we show the individual event rates for sample users 5 and 81 as well as the event rates averaged over all users. The overall behavior of the individual and user-averaged event rates reflects typical weekly cycles of humans by being more active in the daytime and on weekdays and less active in the nighttime and on weekends. From the user-averaged event rates, we find that email (call) is more used around noon (late afternoon) on weekdays, while email (call) is less (more) used than other services in the weekend daytime. Since most users in our dataset were students and staff members of the university, they might not be making or receiving calls in classes or laboratories in the weekday daytime. Instead they might be using other communication services, such as email and SMS. On the other hand, users might be using call more than email outside class or laboratory on weekends. 

To investigate the temporal correlations between service usages for each user, we calculate the Pearson correlation coefficient (PCC) by using the event rates of services $s$ and $s'$ for user $i$:
\begin{equation}
    r_{i,ss'}=\frac{\sum_t [\rho_{is}(t)-\bar \rho_{is}][\rho_{is'}(t)-\bar \rho_{is'}]}
    {\sqrt{\sum_t [\rho_{is}(t)-\bar \rho_{is}]^2}\sqrt{\sum_t [\rho_{is'}(t)-\bar \rho_{is'}]^2}},
    \label{eq:PCC}
\end{equation}
where $\bar \rho_{is}=T_d^{-1}\sum_t \rho_{is}(t)$. For the PCC on weekdays and on weekends, $\rho^{\rm wd}_{is}(t)$ and $\rho^{\rm we}_{is}(t)$ are used, respectively. The values of PCC turn out in most cases to be positive (not shown here). This is mainly due to the typical weekly cycles of humans as mentioned before. To correct such cycles, for each case of weekdays and weekends we consider de-seasoned event rates defined as
\begin{equation}
    \Delta \rho_{is}(t)=\rho_{is}(t)-\frac{1}{S_i}\sum_s \rho_{is}(t),
\end{equation}
where $S_i$ denotes the number of services the user $i$ have used.

As shown in Fig.~\ref{fig:actWeekCorrel}, the values of PCC obtained for the de-seasoned event rates show similar and distinct behavior among users as well as between weekdays and weekends. For example, in the case of user 5, the strongly positive correlation between call and SMS usages on weekdays turns to be slightly negative on weekends. This result is consistent with the temporal patterns depicted in Fig.~\ref{fig:actShortPattern}. The positive (negative) correlation between services by being used at the same time (at different times) of the week can be interpreted such that those services are complementary (substitutive) with each other~\cite{Karikoski2011c}. Then, we obtain and compare distributions of PCC over all users for each pair of services. The mean values for web-app and call-SMS pairs (app-email pair) are slightly positive (negative) on weekdays and become slightly negative (positive) on weekends. All other pairs have the negative mean values. The result for positive correlations is inconclusive due to the large standard errors of PCC up to $0.05$. However, for the pairs of services with large negative correlations, such as web-call and web-SMS pairs, we can argue that those services might be used in a substitutive way. In order to compare the correlations for weekdays and for weekends, we have conducted the Kolmogorov-Smirnov test. It is found that the distributions of PCC for weekdays and for weekends are significantly different for the pairs of web-app ($p$-value less than $0.005$), app-email ($0.03$), email-call ($0.03$), email-SMS ($0.03$), and call-SMS ($0.02$). This list of pairs contains all the pairs whose sign of the mean has changed from weekdays to weekends.

For more detailed, i.e. event-based analysis of correlations among service usages, we obtain the distribution of time interval between two consecutive or simultaneous events but of different services of the same user. Precisely, the time interval for a pair of services $s$ and $s'$ is defined by $\Delta t_{ss'}=t_{s'}-t_s$ with event timings $t_s$ and $t_s'$. As shown in the upper panels of Fig.~\ref{fig:crossCorrel}~(a), distributions for some service pairs have a peak at the negative value of $\Delta t_{ss'}$ both for weekdays and for weekends. This indicates that the event of service $s$ follows that of service $s'$. On the other hand, distributions for other pairs of services do not show any distinct peaks, implying no temporal correlation. This time-ordering behavior could mean that one service usage might effectively induce another service usage. However, we cannot investigate such a process by our dataset. We summarize the results such that communication services, such as email, call and SMS, are followed by non-communication services, i.e. web and app, as depicted in Fig.~\ref{fig:crossCorrel}~(b). We also obtain the distributions of time interval for different contexts. We find the overall similar time-ordering behavior (not shown here), except that email is followed by web at Home and that app does not follow communication services abroad. Note that the event-based analysis cannot be directly compared to the analysis of aggregated weekly patterns.

\begin{figure}[!t]
\includegraphics[width=\columnwidth]{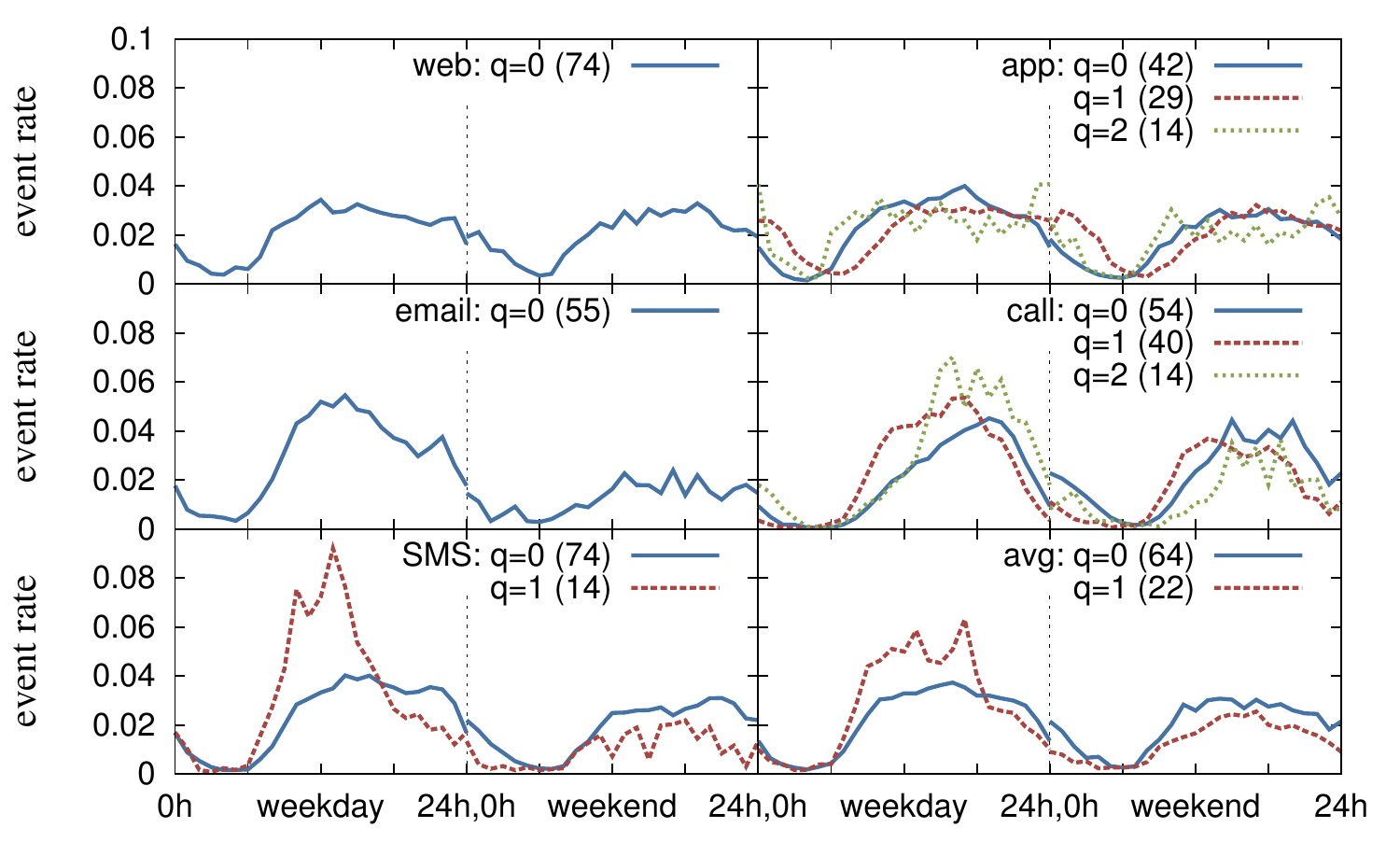}
\caption{$k$-means clustering results of users' weekly patterns. We have used $k=10$ and plotted only a few dominant clusters with cluster size in the parenthesis, see Table~\ref{table:kmeansPartition} for details. The bin size was set to one hour.}
\label{fig:kmeansCluster}
\end{figure}

\begin{figure}[!t]
\includegraphics[width=\columnwidth]{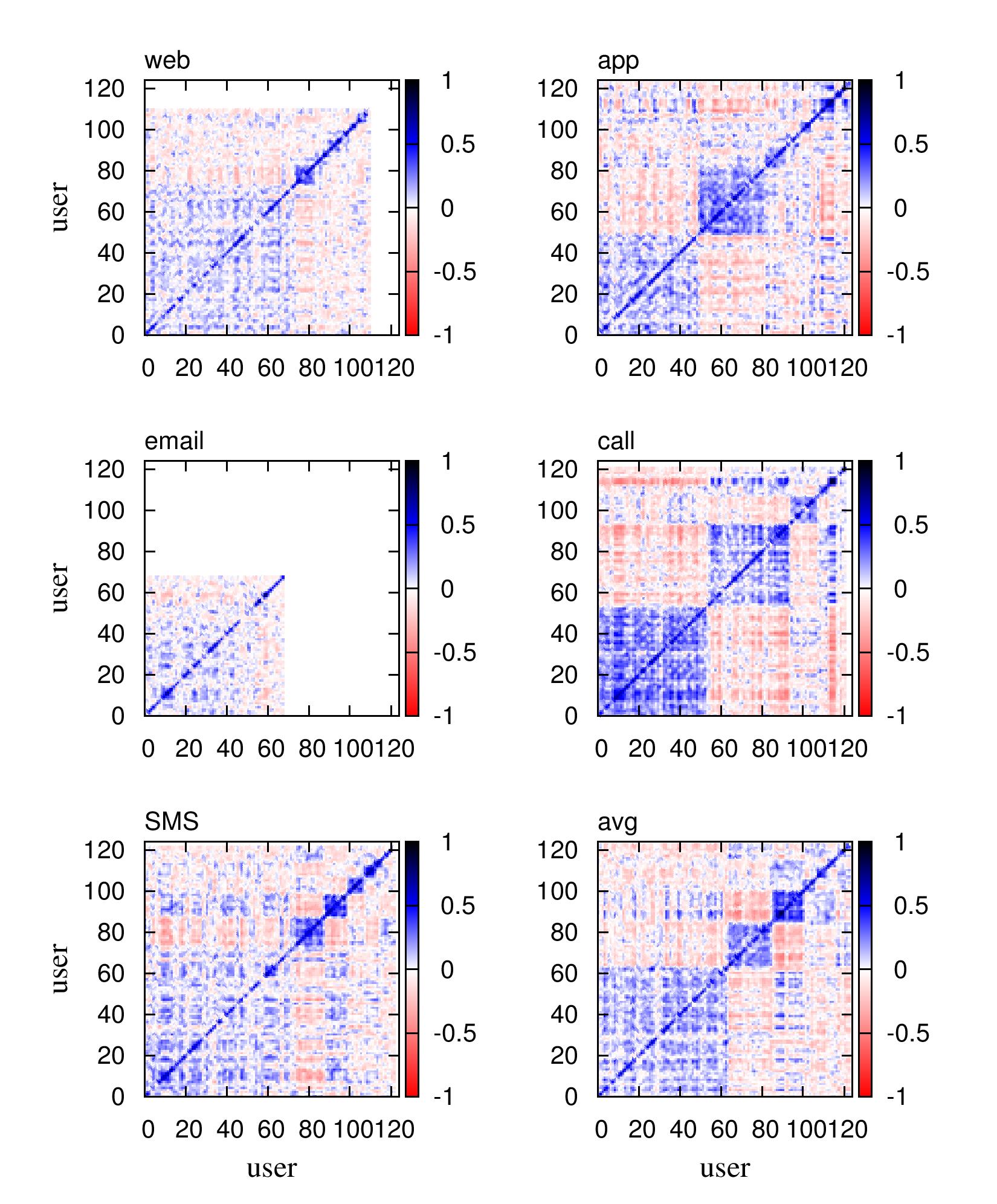}
\caption{Pearson correlation matrices of users' de-seasoned event rates. These matrices support the validity of the $k$-means clustering results in Table~\ref{table:kmeansPartition}. The user index has been sorted according to the corresponding cluster index and blank spaces are due to totally inactive users.}
\label{fig:kmeansCorrelMatrix}\end{figure}

\subsection{Clustering and overlaps in temporal patterns of service usage}

As it turns out, the temporal patterns of service usage are diverse from one user to another, while some of them still show similar behavior. To investigate the similarity and diversity of weekly patterns for each service we apply the $k$-means clustering method~\cite{Gan2007} to the weekly event rates as $\rho_{is}(t)\equiv\{\rho^{\rm wd}_{is}(t),\rho^{\rm we}_{is}(t)\}$. To correct the typical weekly cycles of each service (not of each user), we use the de-seasoned event rates as follows
\begin{eqnarray}
    \Delta \rho_{is}(t)=\rho_{is}(t)-\frac{1}{N_s}\sum_{i} \rho_{is}(t),
\end{eqnarray}
where $N_s$ denotes the number of users showing any activity in service $s$. We similarly define the service-averaged event rates for each user for the clustering, to be denoted by avg. In each case we set the number of clusters as $k=10$ and the cluster index is denoted by $q=0,\cdots,9$. Clustering has been conducted $2000$ times with different initial conditions and here we present the result maximizing the quality of clustering or validity index, defined as the minimum inter-cluster distance divided by the sum of intra-cluster distances~\cite{Gan2007}.

\begin{table}[!b]
\caption{$k$-means clustering results for weekly patterns of service usages with $k=10$. $q$ and $N_s$ denote the cluster index and the number of available users for service $s$, respectively.}
\label{table:kmeansPartition}
\begin{tabular}{lrrrrrrrrrrr}
\hline
service & $q=0$ & $\ 1$ & $\ 2$ & $\ 3$ & $\ 4$ & $\ 5$ & $\ 6$ & $\ 7$ & $\ 8$ & $\ 9$ & $N_s$\\
\hline
web & 74 & 9 & 7 & 6 & 5 & 3 & 3 & 2 & 1 & 1 & 111 \\
app & 50 & 32 & 10 & 7 & 6 & 6 & 5 & 4 & 3 & 1 & 124 \\
email & 55 & 3 & 3 & 2 & 1 & 1 & 1 & 1 & 1 & 1 & 69 \\
call & 54 & 40 & 14 & 5 & 4 & 1 & 1 & 1 & 1 & 1 & 122 \\
SMS & 74 & 14 & 11 & 9 & 5 & 4 & 3 & 1 & 1 & 1 & 123 \\
avg & 64 & 21 & 16 & 6 & 5 & 5 & 4 & 1 & 1 & 1 & 124 \\
\hline
\end{tabular}
\end{table}

The clustering results are summarized in Table~\ref{table:kmeansPartition} and only a few weekly patterns of dominant clusters are shown in Fig.~\ref{fig:kmeansCluster}. Only one dominant cluster is found in each case of web and email usages, implying similar patterns among users. Weekly patterns of app, call, and SMS usages are clustered into more than one dominant cluster. Compared to the largest cluster ($q=0$) of call usage, the second largest cluster ($q=1$) can be characterized by larger activities in the weekday daytime and in the weekend morning. The behavioral difference between dominant clusters in SMS usage is also obvious. The largest cluster ($q=0$) represents the evening-type users, while the second largest cluster ($q=1$) does the morning-type users on weekdays. In the case of service-averaged usage patterns, the second largest cluster ($q=1$) shows the larger (smaller) activity in the daytime on weekdays (on weekends) than the largest cluster ($q=0$). To check the validity of clustering results, we obtain the Pearson correlation matrices using the de-seasoned event rates, $\Delta\rho_{is}(t)$. All the matrices support the $k$-means clustering results, see Fig.~\ref{fig:kmeansCorrelMatrix}. We also tested the effect of the number of means, $k$, on the clustering and found that the results are qualitatively similar apart from the number of small or outlying clusters.

\begin{figure}[!t]
\includegraphics[width=\columnwidth]{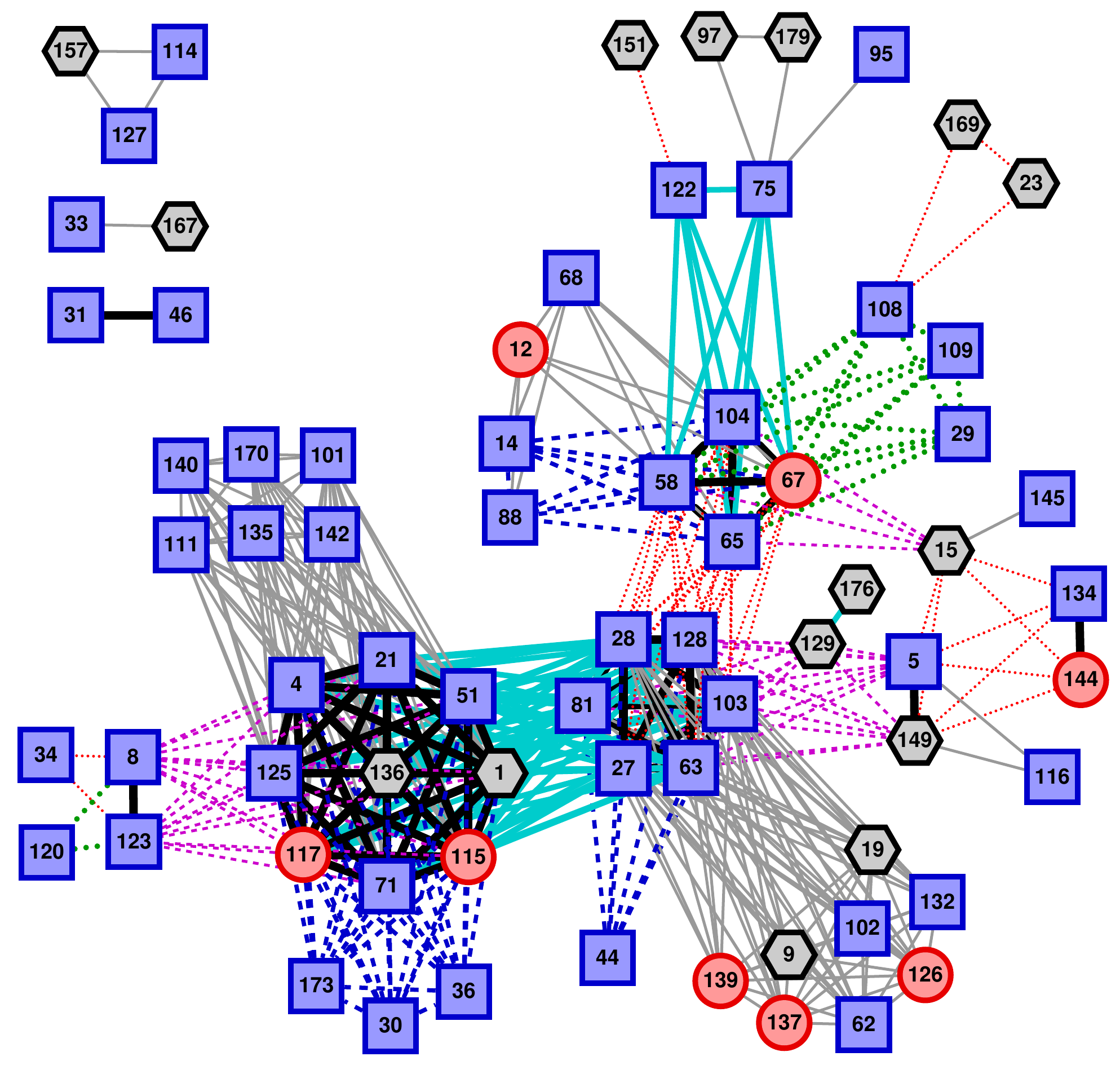}
\caption{Overlap network constructed based on the clustering results for all services. Circle, square, and hexagonal nodes represent female, male, and unknown gender of users, respectively. Each black solid thick line denotes a link between users who belong to the same clusters for all services. Other colored lines denote the links between users who belong to the same clusters for all but one service: web (dashed thick blue), app (dotted thin red), email (dotted thick green), call (solid thick cyan), SMS (dashed thin violet), or due to the unused service by either user (solid thin gray). This figure was generated using Cytoscape v2.8.1~\cite{Smoot2011}.}
\label{fig:overlapNet}\end{figure}

Finally, in order to get insight into the overall structure of temporal correlations among users and services, we construct an overlap network based on the clustering results. This leads to the network of overlapping communities~\cite{Palla2005}, where nodes and link weights of the network represent users and their overlaps, respectively. Precisely, the behavioral overlap is defined as the number of services in which two users, say $i$ and $j$, belong to the same cluster as
\begin{equation}
    O^{\rm B}_{ij}=\sum_{s}\delta(q_{is},q_{js}).
\end{equation}
Here $q_{is}$ denotes a cluster index for user $i$'s service $s$, and the Kronecker delta function $\delta(q,q')$ gives $1$ if $q=q'$ and $0$ otherwise. Figure~\ref{fig:overlapNet} shows the overlap network with 436 links of $O^{\rm B}=4$ and $5$. The behavioral overlap $O^{\rm B}=5$ of a link, denoted by thick black line, implies that the neighboring users belong to the same clusters for all services, i.e. they are fully synchronized. We find cliques consisting of only the fully synchronized users, which we call synchronized cores. The largest synchronized core with 9 users is closely related to the second largest synchronized core except for belonging to different clusters of call usage. These cores are also connected to many other users but not as a synchronized core. This agglomerate structure can be induced by the relatively homogeneous demographics of users in our dataset. However, we like to note that the clustering was applied to the de-seasoned event rates, which have been subtracted by the user-averaged temporal behavior. 

\begin{figure}[!t]
\includegraphics[width=.9\columnwidth]{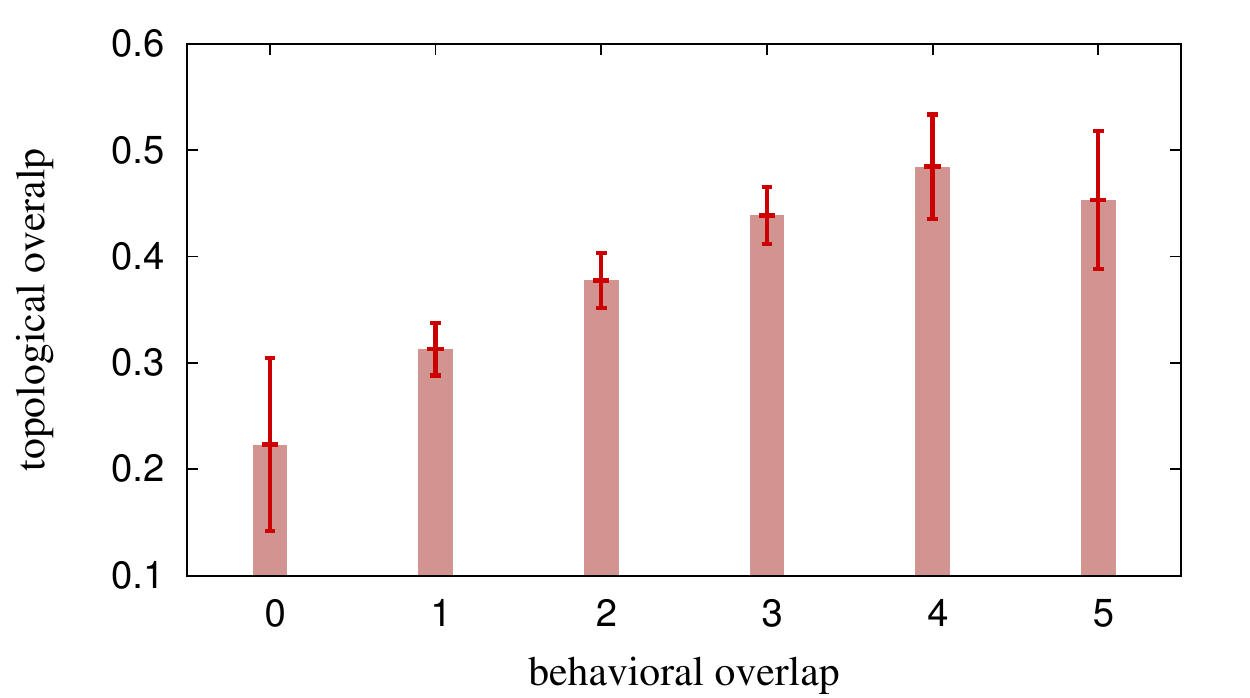}
\caption{Topological overlap as a function of behavioral overlap. The overall positive correlation between two overlaps is observed.}
\label{fig:overlapCorrel}\end{figure}

We compare the behavioral overlap network based on the clustering results to the communication network of users. The communication network can be constructed from the call and SMS datasets containing the information on communication partners. Only 67 out of 124 users and 205 links between users are identified. The topological overlap of a link $ij$ is defined as~\cite{Onnela2007a}
\begin{equation}
    O^{\rm T}_{ij}=\frac{|\Lambda_i \cap \Lambda_j|}{|\Lambda_i \cup \Lambda_j|-2},
\end{equation}
where $\Lambda_i$ denotes the set of neighbors of node $i$. $O^{\rm T}_{ij}$ has a value of 1 if $i$ and $j$ have exactly the same neighbors except for themselves and it has a value of 0 if they do not have any neighbors in common. Figure~\ref{fig:overlapCorrel} shows the overall positive correlation between behavioral and topological overlaps. It implies that connected users sharing more common neighbors show more similar weekly patterns of service usages. Thus, the behavioral overlap network based on the service usages can be used to reveal the communication network structure of users.

\section{Summary}
\label{sec:summary}

We have investigated spatiotemporal correlations and temporal diversities of service usages by analyzing a handset-based dataset collected from 124 users for over 16 months. The dataset consists of locations and service usages. After constructing the precise spatiotemporal trajectory for each user based on the location dataset, we identify several meaningful places or contexts by means of context detection method. As contexts, Home, Office, Other meaningful place, Elsewhere, and Abroad are considered. We showed how the context affects the service usage patterns of users, including their web domain visit (web), application (app), email, voice call (call), and short message service (SMS). 

In this study we have found the similarity and diversity of weekly patterns among users and services, in terms of temporal correlations, time-ordering behavior between services, and overlap network based on clustering. The services used at the same time (at different times) of the week lead to the positive (negative) correlations between them, which can be interpreted as being complementary (substitutive) to each other. By conducting the event-based analysis instead of weekly patterns we observe the time-ordering behavior between services, such that communication services, i.e. email, call, and SMS, are followed by the non-communication services, i.e. web and app. Finally, the similarity and diversity of weekly patterns of service usages enable us to classify users into several different clusters, e.g. as characterized by the morning-type or evening-type usage patterns, except for the web and email usages. The behavioral overlap network constructed based on the clustering results can be used to reveal the communication or real social network structure of users. 

Our findings on the spatiotemporal correlations of service usage patterns for different contexts enable us to better understand the behavior of humans and what that implies. This is also important for better design of information and communications technology (ICT) enabled social environments and services. However, more detailed analysis with higher resolution is required to reveal the underlying mechanism or the origin of spatiotemporal correlations.

\acknowledgments
The research data were collected in the OtaSizzle project that is funded by Aalto University's MIDE program and Helsinki University of Technology TKK's `Technology for Life' campaign donations from private companies and communities. The authors thank MobiTrack Innovations Ltd. for providing the mobile audience measurement platform. The sponsoring from Nokia and Elisa to this work is also acknowledged. Financial support by Aalto University postdoctoral program (HJ), from EU's 7th Framework Program's FET-Open to ICTeCollective project no. 238597, by the Academy of Finland, the Finnish Center of Excellence program 2006-2011, project no. 129670 (MK, KK), and by Future Internet Graduate School and MoMIE project (JK) are gratefully acknowledged.

\bibliographystyle{apsrev4-1}

\begin{thebibliography}{41}%
\makeatletter
\providecommand \@ifxundefined [1]{%
 \@ifx{#1\undefined}
}%
\providecommand \@ifnum [1]{%
 \ifnum #1\expandafter \@firstoftwo
 \else \expandafter \@secondoftwo
 \fi
}%
\providecommand \@ifx [1]{%
 \ifx #1\expandafter \@firstoftwo
 \else \expandafter \@secondoftwo
 \fi
}%
\providecommand \natexlab [1]{#1}%
\providecommand \enquote  [1]{``#1''}%
\providecommand \bibnamefont  [1]{#1}%
\providecommand \bibfnamefont [1]{#1}%
\providecommand \citenamefont [1]{#1}%
\providecommand \href@noop [0]{\@secondoftwo}%
\providecommand \href [0]{\begingroup \@sanitize@url \@href}%
\providecommand \@href[1]{\@@startlink{#1}\@@href}%
\providecommand \@@href[1]{\endgroup#1\@@endlink}%
\providecommand \@sanitize@url [0]{\catcode `\\12\catcode `\$12\catcode
  `\&12\catcode `\#12\catcode `\^12\catcode `\_12\catcode `\%12\relax}%
\providecommand \@@startlink[1]{}%
\providecommand \@@endlink[0]{}%
\providecommand \url  [0]{\begingroup\@sanitize@url \@url }%
\providecommand \@url [1]{\endgroup\@href {#1}{\urlprefix }}%
\providecommand \urlprefix  [0]{URL }%
\providecommand \Eprint [0]{\href }%
\providecommand \doibase [0]{http://dx.doi.org/}%
\providecommand \selectlanguage [0]{\@gobble}%
\providecommand \bibinfo  [0]{\@secondoftwo}%
\providecommand \bibfield  [0]{\@secondoftwo}%
\providecommand \translation [1]{[#1]}%
\providecommand \BibitemOpen [0]{}%
\providecommand \bibitemStop [0]{}%
\providecommand \bibitemNoStop [0]{.\EOS\space}%
\providecommand \EOS [0]{\spacefactor3000\relax}%
\providecommand \BibitemShut  [1]{\csname bibitem#1\endcsname}%
\let\auto@bib@innerbib\@empty
\bibitem [{\citenamefont {Goyal}(2009)}]{Goyal2009}%
  \BibitemOpen
  \bibfield  {author} {\bibinfo {author} {\bibfnamefont {S.}~\bibnamefont
  {Goyal}},\ }\href@noop {} {\emph {\bibinfo {title} {Connections : an
  introduction to the network economy}}}\ (\bibinfo  {publisher} {Princeton
  University Press},\ \bibinfo {year} {2009})\BibitemShut {NoStop}%
\bibitem [{\citenamefont {Castellano}\ \emph {et~al.}(2009)\citenamefont
  {Castellano}, \citenamefont {Fortunato},\ and\ \citenamefont
  {Loreto}}]{Castellano2009}%
  \BibitemOpen
  \bibfield  {author} {\bibinfo {author} {\bibfnamefont {C.}~\bibnamefont
  {Castellano}}, \bibinfo {author} {\bibfnamefont {S.}~\bibnamefont
  {Fortunato}}, \ and\ \bibinfo {author} {\bibfnamefont {V.}~\bibnamefont
  {Loreto}},\ }\href {\doibase 10.1103/RevModPhys.81.591} {\bibfield  {journal}
  {\bibinfo  {journal} {Reviews of Modern Physics}\ }\textbf {\bibinfo {volume}
  {81}},\ \bibinfo {pages} {591} (\bibinfo {year} {2009})}\BibitemShut
  {NoStop}%
\bibitem [{\citenamefont {Lazer}\ \emph {et~al.}(2009)\citenamefont {Lazer},
  \citenamefont {Pentland}, \citenamefont {Adamic}, \citenamefont {Aral},
  \citenamefont {Barab\'asi}, \citenamefont {Brewer}, \citenamefont
  {Christakis}, \citenamefont {Contractor}, \citenamefont {Fowler},
  \citenamefont {Gutmann}, \citenamefont {Jebara}, \citenamefont {King},
  \citenamefont {Macy}, \citenamefont {Roy},\ and\ \citenamefont
  {Van~Alstyne}}]{Lazer2009}%
  \BibitemOpen
  \bibfield  {author} {\bibinfo {author} {\bibfnamefont {D.}~\bibnamefont
  {Lazer}}, \bibinfo {author} {\bibfnamefont {A.}~\bibnamefont {Pentland}},
  \bibinfo {author} {\bibfnamefont {L.}~\bibnamefont {Adamic}}, \bibinfo
  {author} {\bibfnamefont {S.}~\bibnamefont {Aral}}, \bibinfo {author}
  {\bibfnamefont {A.-L.}\ \bibnamefont {Barab\'asi}}, \bibinfo {author}
  {\bibfnamefont {D.}~\bibnamefont {Brewer}}, \bibinfo {author} {\bibfnamefont
  {N.}~\bibnamefont {Christakis}}, \bibinfo {author} {\bibfnamefont
  {N.}~\bibnamefont {Contractor}}, \bibinfo {author} {\bibfnamefont
  {J.}~\bibnamefont {Fowler}}, \bibinfo {author} {\bibfnamefont
  {M.}~\bibnamefont {Gutmann}}, \bibinfo {author} {\bibfnamefont
  {T.}~\bibnamefont {Jebara}}, \bibinfo {author} {\bibfnamefont
  {G.}~\bibnamefont {King}}, \bibinfo {author} {\bibfnamefont {M.}~\bibnamefont
  {Macy}}, \bibinfo {author} {\bibfnamefont {D.}~\bibnamefont {Roy}}, \ and\
  \bibinfo {author} {\bibfnamefont {M.}~\bibnamefont {Van~Alstyne}},\ }\href
  {\doibase 10.1126/science.1167742} {\bibfield  {journal} {\bibinfo  {journal}
  {Science}\ }\textbf {\bibinfo {volume} {323}},\ \bibinfo {pages} {721}
  (\bibinfo {year} {2009})}\BibitemShut {NoStop}%
\bibitem [{\citenamefont {Eckmann}\ \emph {et~al.}(2004)\citenamefont
  {Eckmann}, \citenamefont {Moses},\ and\ \citenamefont {Sergi}}]{Eckmann2004}%
  \BibitemOpen
  \bibfield  {author} {\bibinfo {author} {\bibfnamefont {J.-P.}\ \bibnamefont
  {Eckmann}}, \bibinfo {author} {\bibfnamefont {E.}~\bibnamefont {Moses}}, \
  and\ \bibinfo {author} {\bibfnamefont {D.}~\bibnamefont {Sergi}},\ }\href
  {\doibase 10.1073/pnas.0405728101} {\bibfield  {journal} {\bibinfo  {journal}
  {Proceedings of the National Academy of Sciences}\ }\textbf {\bibinfo
  {volume} {101}},\ \bibinfo {pages} {14333} (\bibinfo {year}
  {2004})}\BibitemShut {NoStop}%
\bibitem [{\citenamefont {Barab\'{a}si}(2005)}]{Barabasi2005}%
  \BibitemOpen
  \bibfield  {author} {\bibinfo {author} {\bibfnamefont {A.-L.}\ \bibnamefont
  {Barab\'{a}si}},\ }\href {\doibase 10.1038/nature03459} {\bibfield  {journal}
  {\bibinfo  {journal} {Nature}\ }\textbf {\bibinfo {volume} {435}},\ \bibinfo
  {pages} {207} (\bibinfo {year} {2005})}\BibitemShut {NoStop}%
\bibitem [{\citenamefont {Onnela}\ \emph {et~al.}(2007)\citenamefont {Onnela},
  \citenamefont {Saram\"{a}ki}, \citenamefont {Hyv\"{o}nen}, \citenamefont
  {Szab\'{o}}, \citenamefont {Lazer}, \citenamefont {Kaski}, \citenamefont
  {Kert\'{e}sz},\ and\ \citenamefont {Barab\'{a}si}}]{Onnela2007a}%
  \BibitemOpen
  \bibfield  {author} {\bibinfo {author} {\bibfnamefont {J.~P.}\ \bibnamefont
  {Onnela}}, \bibinfo {author} {\bibfnamefont {J.}~\bibnamefont
  {Saram\"{a}ki}}, \bibinfo {author} {\bibfnamefont {J.}~\bibnamefont
  {Hyv\"{o}nen}}, \bibinfo {author} {\bibfnamefont {G.}~\bibnamefont
  {Szab\'{o}}}, \bibinfo {author} {\bibfnamefont {D.}~\bibnamefont {Lazer}},
  \bibinfo {author} {\bibfnamefont {K.}~\bibnamefont {Kaski}}, \bibinfo
  {author} {\bibfnamefont {J.}~\bibnamefont {Kert\'{e}sz}}, \ and\ \bibinfo
  {author} {\bibfnamefont {A.~L.}\ \bibnamefont {Barab\'{a}si}},\ }\href
  {\doibase 10.1073/pnas.0610245104} {\bibfield  {journal} {\bibinfo  {journal}
  {Proceedings of the National Academy of Sciences}\ }\textbf {\bibinfo
  {volume} {104}},\ \bibinfo {pages} {7332} (\bibinfo {year}
  {2007})}\BibitemShut {NoStop}%
\bibitem [{\citenamefont {Kwak}\ \emph {et~al.}(2010)\citenamefont {Kwak},
  \citenamefont {Lee}, \citenamefont {Park},\ and\ \citenamefont
  {Moon}}]{Kwak2010}%
  \BibitemOpen
  \bibfield  {author} {\bibinfo {author} {\bibfnamefont {H.}~\bibnamefont
  {Kwak}}, \bibinfo {author} {\bibfnamefont {C.}~\bibnamefont {Lee}}, \bibinfo
  {author} {\bibfnamefont {H.}~\bibnamefont {Park}}, \ and\ \bibinfo {author}
  {\bibfnamefont {S.}~\bibnamefont {Moon}},\ }in\ \href {\doibase
  10.1145/1772690.1772751} {\emph {\bibinfo {booktitle} {Proceedings of the
  19th international conference on World wide web}}},\ \bibinfo {series and
  number} {WWW '10}\ (\bibinfo  {publisher} {ACM},\ \bibinfo {address} {New
  York, NY, USA},\ \bibinfo {year} {2010})\ pp.\ \bibinfo {pages}
  {591--600}\BibitemShut {NoStop}%
\bibitem [{\citenamefont {Lewis}\ \emph {et~al.}(2008)\citenamefont {Lewis},
  \citenamefont {Kaufman}, \citenamefont {Gonzalez}, \citenamefont {Wimmer},\
  and\ \citenamefont {Christakis}}]{Lewis2008}%
  \BibitemOpen
  \bibfield  {author} {\bibinfo {author} {\bibfnamefont {K.}~\bibnamefont
  {Lewis}}, \bibinfo {author} {\bibfnamefont {J.}~\bibnamefont {Kaufman}},
  \bibinfo {author} {\bibfnamefont {M.}~\bibnamefont {Gonzalez}}, \bibinfo
  {author} {\bibfnamefont {A.}~\bibnamefont {Wimmer}}, \ and\ \bibinfo {author}
  {\bibfnamefont {N.}~\bibnamefont {Christakis}},\ }\href {\doibase
  10.1016/j.socnet.2008.07.002} {\bibfield  {journal} {\bibinfo  {journal}
  {Social Networks}\ }\textbf {\bibinfo {volume} {30}},\ \bibinfo {pages} {330}
  (\bibinfo {year} {2008})}\BibitemShut {NoStop}%
\bibitem [{\citenamefont {Kovanen}\ \emph {et~al.}(2011)\citenamefont
  {Kovanen}, \citenamefont {Karsai}, \citenamefont {Kaski}, \citenamefont
  {Kert\'{e}sz},\ and\ \citenamefont {Saram\"{a}ki}}]{Kovanen2011}%
  \BibitemOpen
  \bibfield  {author} {\bibinfo {author} {\bibfnamefont {L.}~\bibnamefont
  {Kovanen}}, \bibinfo {author} {\bibfnamefont {M.}~\bibnamefont {Karsai}},
  \bibinfo {author} {\bibfnamefont {K.}~\bibnamefont {Kaski}}, \bibinfo
  {author} {\bibfnamefont {J.}~\bibnamefont {Kert\'{e}sz}}, \ and\ \bibinfo
  {author} {\bibfnamefont {J.}~\bibnamefont {Saram\"{a}ki}},\ }\href {\doibase
  10.1088/1742-5468/2011/11/P11005} {\bibfield  {journal} {\bibinfo  {journal}
  {Journal of Statistical Mechanics: Theory and Experiment}\ }\textbf {\bibinfo
  {volume} {2011}},\ \bibinfo {pages} {P11005} (\bibinfo {year}
  {2011})}\BibitemShut {NoStop}%
\bibitem [{\citenamefont {Jo}\ \emph {et~al.}(2012)\citenamefont {Jo},
  \citenamefont {Karsai}, \citenamefont {Kert\'{e}sz},\ and\ \citenamefont
  {Kaski}}]{Jo2012}%
  \BibitemOpen
  \bibfield  {author} {\bibinfo {author} {\bibfnamefont {H.-H.}\ \bibnamefont
  {Jo}}, \bibinfo {author} {\bibfnamefont {M.}~\bibnamefont {Karsai}}, \bibinfo
  {author} {\bibfnamefont {J.}~\bibnamefont {Kert\'{e}sz}}, \ and\ \bibinfo
  {author} {\bibfnamefont {K.}~\bibnamefont {Kaski}},\ }\href {\doibase
  10.1088/1367-2630/14/1/013055} {\bibfield  {journal} {\bibinfo  {journal}
  {New Journal of Physics}\ }\textbf {\bibinfo {volume} {14}},\ \bibinfo
  {pages} {013055} (\bibinfo {year} {2012})}\BibitemShut {NoStop}%
\bibitem [{\citenamefont {Karsai}\ \emph {et~al.}(2012)\citenamefont {Karsai},
  \citenamefont {Kaski}, \citenamefont {Barab\'{a}si},\ and\ \citenamefont
  {Kert\'{e}sz}}]{Karsai2012}%
  \BibitemOpen
  \bibfield  {author} {\bibinfo {author} {\bibfnamefont {M.}~\bibnamefont
  {Karsai}}, \bibinfo {author} {\bibfnamefont {K.}~\bibnamefont {Kaski}},
  \bibinfo {author} {\bibfnamefont {A.-L.}\ \bibnamefont {Barab\'{a}si}}, \
  and\ \bibinfo {author} {\bibfnamefont {J.}~\bibnamefont {Kert\'{e}sz}},\
  }\href@noop {} {\bibfield  {journal} {\bibinfo  {journal} {Scientific
  Reports}\ }\textbf {\bibinfo {volume} {2}},\ \bibinfo {pages} {397} (\bibinfo
  {year} {2012})}\BibitemShut {NoStop}%
\bibitem [{\citenamefont {Gonz\'{a}lez}\ \emph {et~al.}(2008)\citenamefont
  {Gonz\'{a}lez}, \citenamefont {Hidalgo},\ and\ \citenamefont
  {Barab\'{a}si}}]{Gonzalez2008}%
  \BibitemOpen
  \bibfield  {author} {\bibinfo {author} {\bibfnamefont {M.~C.}\ \bibnamefont
  {Gonz\'{a}lez}}, \bibinfo {author} {\bibfnamefont {C.~A.}\ \bibnamefont
  {Hidalgo}}, \ and\ \bibinfo {author} {\bibfnamefont {A.-L.}\ \bibnamefont
  {Barab\'{a}si}},\ }\href {\doibase 10.1038/nature06958} {\bibfield  {journal}
  {\bibinfo  {journal} {Nature}\ }\textbf {\bibinfo {volume} {453}},\ \bibinfo
  {pages} {779} (\bibinfo {year} {2008})}\BibitemShut {NoStop}%
\bibitem [{\citenamefont {Candia}\ \emph {et~al.}(2008)\citenamefont {Candia},
  \citenamefont {Gonz\'{a}lez}, \citenamefont {Wang}, \citenamefont
  {Schoenharl}, \citenamefont {Madey},\ and\ \citenamefont
  {Barab\'{a}si}}]{Candia2008}%
  \BibitemOpen
  \bibfield  {author} {\bibinfo {author} {\bibfnamefont {J.}~\bibnamefont
  {Candia}}, \bibinfo {author} {\bibfnamefont {M.~C.}\ \bibnamefont
  {Gonz\'{a}lez}}, \bibinfo {author} {\bibfnamefont {P.}~\bibnamefont {Wang}},
  \bibinfo {author} {\bibfnamefont {T.}~\bibnamefont {Schoenharl}}, \bibinfo
  {author} {\bibfnamefont {G.}~\bibnamefont {Madey}}, \ and\ \bibinfo {author}
  {\bibfnamefont {A.-L.}\ \bibnamefont {Barab\'{a}si}},\ }\href {\doibase
  10.1088/1751-8113/41/22/224015} {\bibfield  {journal} {\bibinfo  {journal}
  {Journal of Physics A: Mathematical and Theoretical}\ }\textbf {\bibinfo
  {volume} {41}},\ \bibinfo {pages} {224015} (\bibinfo {year}
  {2008})}\BibitemShut {NoStop}%
\bibitem [{\citenamefont {Wang}\ \emph {et~al.}(2009)\citenamefont {Wang},
  \citenamefont {Gonz\'{a}lez}, \citenamefont {Hidalgo},\ and\ \citenamefont
  {Barab\'{a}si}}]{Wang2009}%
  \BibitemOpen
  \bibfield  {author} {\bibinfo {author} {\bibfnamefont {P.}~\bibnamefont
  {Wang}}, \bibinfo {author} {\bibfnamefont {M.~C.}\ \bibnamefont
  {Gonz\'{a}lez}}, \bibinfo {author} {\bibfnamefont {C.~A.}\ \bibnamefont
  {Hidalgo}}, \ and\ \bibinfo {author} {\bibfnamefont {A.-L.}\ \bibnamefont
  {Barab\'{a}si}},\ }\href {\doibase 10.1126/science.1167053} {\bibfield
  {journal} {\bibinfo  {journal} {Science}\ }\textbf {\bibinfo {volume}
  {324}},\ \bibinfo {pages} {1071} (\bibinfo {year} {2009})}\BibitemShut
  {NoStop}%
\bibitem [{\citenamefont {Song}\ \emph
  {et~al.}(2010{\natexlab{a}})\citenamefont {Song}, \citenamefont {Qu},
  \citenamefont {Blumm},\ and\ \citenamefont {Barab\'{a}si}}]{Song2010a}%
  \BibitemOpen
  \bibfield  {author} {\bibinfo {author} {\bibfnamefont {C.}~\bibnamefont
  {Song}}, \bibinfo {author} {\bibfnamefont {Z.}~\bibnamefont {Qu}}, \bibinfo
  {author} {\bibfnamefont {N.}~\bibnamefont {Blumm}}, \ and\ \bibinfo {author}
  {\bibfnamefont {A.-L.}\ \bibnamefont {Barab\'{a}si}},\ }\href {\doibase
  10.1126/science.1177170} {\bibfield  {journal} {\bibinfo  {journal}
  {Science}\ }\textbf {\bibinfo {volume} {327}},\ \bibinfo {pages} {1018}
  (\bibinfo {year} {2010}{\natexlab{a}})}\BibitemShut {NoStop}%
\bibitem [{\citenamefont {Song}\ \emph
  {et~al.}(2010{\natexlab{b}})\citenamefont {Song}, \citenamefont {Koren},
  \citenamefont {Wang},\ and\ \citenamefont {Barabasi}}]{Song2010b}%
  \BibitemOpen
  \bibfield  {author} {\bibinfo {author} {\bibfnamefont {C.}~\bibnamefont
  {Song}}, \bibinfo {author} {\bibfnamefont {T.}~\bibnamefont {Koren}},
  \bibinfo {author} {\bibfnamefont {P.}~\bibnamefont {Wang}}, \ and\ \bibinfo
  {author} {\bibfnamefont {A.-L.}\ \bibnamefont {Barabasi}},\ }\href {\doibase
  10.1038/nphys1760} {\bibfield  {journal} {\bibinfo  {journal} {Nature
  Physics}\ }\textbf {\bibinfo {volume} {6}},\ \bibinfo {pages} {818} (\bibinfo
  {year} {2010}{\natexlab{b}})}\BibitemShut {NoStop}%
\bibitem [{\citenamefont {Eagle}\ and\ \citenamefont
  {Pentland}(2006)}]{Eagle2006}%
  \BibitemOpen
  \bibfield  {author} {\bibinfo {author} {\bibfnamefont {N.}~\bibnamefont
  {Eagle}}\ and\ \bibinfo {author} {\bibfnamefont {A.}~\bibnamefont
  {Pentland}},\ }\href {\doibase 10.1007/s00779-005-0046-3} {\bibfield
  {journal} {\bibinfo  {journal} {Personal Ubiquitous Comput.}\ }\textbf
  {\bibinfo {volume} {10}},\ \bibinfo {pages} {255} (\bibinfo {year}
  {2006})}\BibitemShut {NoStop}%
\bibitem [{\citenamefont {Eagle}\ \emph {et~al.}(2009)\citenamefont {Eagle},
  \citenamefont {Pentland},\ and\ \citenamefont {Lazer}}]{Eagle2009}%
  \BibitemOpen
  \bibfield  {author} {\bibinfo {author} {\bibfnamefont {N.}~\bibnamefont
  {Eagle}}, \bibinfo {author} {\bibfnamefont {A.~S.}\ \bibnamefont {Pentland}},
  \ and\ \bibinfo {author} {\bibfnamefont {D.}~\bibnamefont {Lazer}},\ }\href
  {\doibase 10.1073/pnas.0900282106} {\bibfield  {journal} {\bibinfo  {journal}
  {Proceedings of the National Academy of Sciences}\ }\textbf {\bibinfo
  {volume} {106}},\ \bibinfo {pages} {15274} (\bibinfo {year}
  {2009})}\BibitemShut {NoStop}%
\bibitem [{\citenamefont {Krings}\ \emph {et~al.}(2009)\citenamefont {Krings},
  \citenamefont {Calabrese}, \citenamefont {Ratti},\ and\ \citenamefont
  {Blondel}}]{Krings2009}%
  \BibitemOpen
  \bibfield  {author} {\bibinfo {author} {\bibfnamefont {G.}~\bibnamefont
  {Krings}}, \bibinfo {author} {\bibfnamefont {F.}~\bibnamefont {Calabrese}},
  \bibinfo {author} {\bibfnamefont {C.}~\bibnamefont {Ratti}}, \ and\ \bibinfo
  {author} {\bibfnamefont {V.~D.}\ \bibnamefont {Blondel}},\ }\href {\doibase
  10.1088/1742-5468/2009/07/L07003} {\bibfield  {journal} {\bibinfo  {journal}
  {Journal of Statistical Mechanics: Theory and Experiment}\ }\textbf {\bibinfo
  {volume} {2009}},\ \bibinfo {pages} {L07003} (\bibinfo {year}
  {2009})}\BibitemShut {NoStop}%
\bibitem [{\citenamefont {Bagrow}\ and\ \citenamefont
  {Lin}(2012)}]{Bagrow2012}%
  \BibitemOpen
  \bibfield  {author} {\bibinfo {author} {\bibfnamefont {J.~P.}\ \bibnamefont
  {Bagrow}}\ and\ \bibinfo {author} {\bibfnamefont {Y.-R.}\ \bibnamefont
  {Lin}},\ }\href {\doibase 10.1371/journal.pone.0037676} {\bibfield  {journal}
  {\bibinfo  {journal} {PLoS ONE}\ }\textbf {\bibinfo {volume} {7}},\ \bibinfo
  {pages} {e37676} (\bibinfo {year} {2012})}\BibitemShut {NoStop}%
\bibitem [{\citenamefont {Aharony}\ \emph {et~al.}(2011)\citenamefont
  {Aharony}, \citenamefont {Pan}, \citenamefont {Ip}, \citenamefont {Khayal},\
  and\ \citenamefont {Pentland}}]{Aharony2011}%
  \BibitemOpen
  \bibfield  {author} {\bibinfo {author} {\bibfnamefont {N.}~\bibnamefont
  {Aharony}}, \bibinfo {author} {\bibfnamefont {W.}~\bibnamefont {Pan}},
  \bibinfo {author} {\bibfnamefont {C.}~\bibnamefont {Ip}}, \bibinfo {author}
  {\bibfnamefont {I.}~\bibnamefont {Khayal}}, \ and\ \bibinfo {author}
  {\bibfnamefont {A.}~\bibnamefont {Pentland}},\ }\href {\doibase
  10.1016/j.pmcj.2011.09.004} {\bibfield  {journal} {\bibinfo  {journal}
  {Pervasive and Mobile Computing}\ }\textbf {\bibinfo {volume} {7}},\ \bibinfo
  {pages} {643} (\bibinfo {year} {2011})}\BibitemShut {NoStop}%
\bibitem [{\citenamefont {Falaki}\ \emph {et~al.}(2010)\citenamefont {Falaki},
  \citenamefont {Mahajan}, \citenamefont {Kandula}, \citenamefont
  {Lymberopoulos}, \citenamefont {Govindan},\ and\ \citenamefont
  {Estrin}}]{Falaki2010}%
  \BibitemOpen
  \bibfield  {author} {\bibinfo {author} {\bibfnamefont {H.}~\bibnamefont
  {Falaki}}, \bibinfo {author} {\bibfnamefont {R.}~\bibnamefont {Mahajan}},
  \bibinfo {author} {\bibfnamefont {S.}~\bibnamefont {Kandula}}, \bibinfo
  {author} {\bibfnamefont {D.}~\bibnamefont {Lymberopoulos}}, \bibinfo {author}
  {\bibfnamefont {R.}~\bibnamefont {Govindan}}, \ and\ \bibinfo {author}
  {\bibfnamefont {D.}~\bibnamefont {Estrin}},\ }in\ \href {\doibase
  10.1145/1814433.1814453} {\emph {\bibinfo {booktitle} {Proceedings of the 8th
  international conference on Mobile systems, applications, and services}}},\
  \bibinfo {series and number} {MobiSys '10}\ (\bibinfo  {publisher} {ACM},\
  \bibinfo {address} {New York, NY, USA},\ \bibinfo {year} {2010})\ pp.\
  \bibinfo {pages} {179--194}\BibitemShut {NoStop}%
\bibitem [{\citenamefont {Soikkeli}\ \emph {et~al.}(2011)\citenamefont
  {Soikkeli}, \citenamefont {Karikoski},\ and\ \citenamefont
  {Hammainen}}]{Soikkeli2011b}%
  \BibitemOpen
  \bibfield  {author} {\bibinfo {author} {\bibfnamefont {T.}~\bibnamefont
  {Soikkeli}}, \bibinfo {author} {\bibfnamefont {J.}~\bibnamefont {Karikoski}},
  \ and\ \bibinfo {author} {\bibfnamefont {H.}~\bibnamefont {Hammainen}},\ }in\
  \href {\doibase 10.1109/NGMAST.2011.12} {\emph {\bibinfo {booktitle} {Next
  Generation Mobile Applications, Services and Technologies (NGMAST), 2011 5th
  International Conference on}}}\ (\bibinfo  {publisher} {IEEE},\ \bibinfo
  {year} {2011})\ pp.\ \bibinfo {pages} {7--12}\BibitemShut {NoStop}%
\bibitem [{\citenamefont {Dey}(2001)}]{Dey2001}%
  \BibitemOpen
  \bibfield  {author} {\bibinfo {author} {\bibfnamefont {A.~K.}\ \bibnamefont
  {Dey}},\ }\href {\doibase 10.1007/s007790170019} {\bibfield  {journal}
  {\bibinfo  {journal} {Personal and Ubiquitous Computing}\ }\textbf {\bibinfo
  {volume} {5}},\ \bibinfo {pages} {4} (\bibinfo {year} {2001})}\BibitemShut
  {NoStop}%
\bibitem [{\citenamefont {Verkasalo}(2009)}]{Verkasalo2009}%
  \BibitemOpen
  \bibfield  {author} {\bibinfo {author} {\bibfnamefont {H.}~\bibnamefont
  {Verkasalo}},\ }\emph {\bibinfo {title} {Handset-Based Analysis of Mobile
  Service Usage}},\ \href@noop {} {Ph.D. thesis},\ \bibinfo  {school} {Helsinki
  University of Technology}, \bibinfo {address} {Espoo, Finland} (\bibinfo
  {year} {2009})\BibitemShut {NoStop}%
\bibitem [{\citenamefont {Soikkeli}()}]{Soikkeli2011a}%
  \BibitemOpen
  \bibfield  {author} {\bibinfo {author} {\bibfnamefont {T.}~\bibnamefont
  {Soikkeli}},\ }\emph {\bibinfo {title} {The effect of context on smartphone
  usage sessions}},\ \href {http://aalto-fi.academia.edu/TapioSoikkeli/Papers}
  {Master's thesis},\ \bibinfo  {school} {Aalto University}, \bibinfo {address}
  {Espoo, Finland}\BibitemShut {NoStop}%
\bibitem [{\citenamefont {Karikoski}\ and\ \citenamefont
  {Soikkeli}()}]{Karikoski2011a}%
  \BibitemOpen
  \bibfield  {author} {\bibinfo {author} {\bibfnamefont {J.}~\bibnamefont
  {Karikoski}}\ and\ \bibinfo {author} {\bibfnamefont {T.}~\bibnamefont
  {Soikkeli}},\ }\href@noop {} {\bibinfo  {journal} {Personal and Ubiquitous
  Computing (in press)}\ }\BibitemShut {NoStop}%
\bibitem [{siz()}]{sizl}%
  \BibitemOpen
\bibfield  {journal} {  }\href@noop {} {\enquote {\bibinfo {title} {Otasizzle
  project},}\ }\bibinfo {howpublished} {[http://sizl.org]}\BibitemShut
  {NoStop}%
\bibitem [{\citenamefont {Montoliu}\ and\ \citenamefont
  {Perez}(2010)}]{Montoliu2010}%
  \BibitemOpen
  \bibfield  {author} {\bibinfo {author} {\bibfnamefont {R.}~\bibnamefont
  {Montoliu}}\ and\ \bibinfo {author} {\bibfnamefont {D.~G.}\ \bibnamefont
  {Perez}},\ }in\ \href {\doibase 10.1145/1899475.1899487} {\emph {\bibinfo
  {booktitle} {Proceedings of the 9th International Conference on Mobile and
  Ubiquitous Multimedia}}},\ \bibinfo {series and number} {MUM '10}\ (\bibinfo
  {publisher} {ACM},\ \bibinfo {address} {New York, NY, USA},\ \bibinfo {year}
  {2010})\BibitemShut {NoStop}%
\bibitem [{\citenamefont {Nurmi}\ and\ \citenamefont
  {Koolwaaij}(2006)}]{Nurmi2006}%
  \BibitemOpen
  \bibfield  {author} {\bibinfo {author} {\bibfnamefont {P.}~\bibnamefont
  {Nurmi}}\ and\ \bibinfo {author} {\bibfnamefont {J.}~\bibnamefont
  {Koolwaaij}},\ }in\ \href {\doibase 10.1109/MOBIQ.2006.340429} {\emph
  {\bibinfo {booktitle} {Mobile and Ubiquitous Systems: Networking and
  Services, 2006 Third Annual International Conference on}}}\ (\bibinfo {year}
  {2006})\ pp.\ \bibinfo {pages} {1--8}\BibitemShut {NoStop}%
\bibitem [{\citenamefont {Eagle}\ and\ \citenamefont
  {Pentland}(2009)}]{Eagle2009b}%
  \BibitemOpen
  \bibfield  {author} {\bibinfo {author} {\bibfnamefont {N.}~\bibnamefont
  {Eagle}}\ and\ \bibinfo {author} {\bibfnamefont {A.~S.}\ \bibnamefont
  {Pentland}},\ }\href {\doibase 10.1007/s00265-009-0739-0} {\bibfield
  {journal} {\bibinfo  {journal} {Behavioral Ecology and Sociobiology}\
  }\textbf {\bibinfo {volume} {63}},\ \bibinfo {pages} {1057} (\bibinfo {year}
  {2009})}\BibitemShut {NoStop}%
\bibitem [{\citenamefont {Reades}\ \emph {et~al.}(2009)\citenamefont {Reades},
  \citenamefont {Calabrese},\ and\ \citenamefont {Ratti}}]{Reades2009}%
  \BibitemOpen
  \bibfield  {author} {\bibinfo {author} {\bibfnamefont {J.}~\bibnamefont
  {Reades}}, \bibinfo {author} {\bibfnamefont {F.}~\bibnamefont {Calabrese}}, \
  and\ \bibinfo {author} {\bibfnamefont {C.}~\bibnamefont {Ratti}},\ }\href
  {\doibase 10.1068/b34133t} {\bibfield  {journal} {\bibinfo  {journal}
  {Environment and Planning B: Planning and Design}\ }\textbf {\bibinfo
  {volume} {36}},\ \bibinfo {pages} {824} (\bibinfo {year} {2009})}\BibitemShut
  {NoStop}%
\bibitem [{\citenamefont {Park}\ \emph {et~al.}(2010)\citenamefont {Park},
  \citenamefont {Lee},\ and\ \citenamefont {Gonz\'{a}lez}}]{Park2010}%
  \BibitemOpen
  \bibfield  {author} {\bibinfo {author} {\bibfnamefont {J.}~\bibnamefont
  {Park}}, \bibinfo {author} {\bibfnamefont {D.-S.}\ \bibnamefont {Lee}}, \
  and\ \bibinfo {author} {\bibfnamefont {M.~C.}\ \bibnamefont {Gonz\'{a}lez}},\
  }\href {\doibase 10.1088/1742-5468/2010/11/P11021} {\bibfield  {journal}
  {\bibinfo  {journal} {Journal of Statistical Mechanics: Theory and
  Experiment}\ }\textbf {\bibinfo {volume} {2010}},\ \bibinfo {pages} {P11021}
  (\bibinfo {year} {2010})}\BibitemShut {NoStop}%
\bibitem [{\citenamefont {Karikoski}()}]{Karikoski2012}%
  \BibitemOpen
  \bibfield  {author} {\bibinfo {author} {\bibfnamefont {J.}~\bibnamefont
  {Karikoski}},\ }\href@noop {} {\bibinfo  {journal} {International Journal of
  Handheld Computing Research (IJHCR) (in press)}\ }\BibitemShut {NoStop}%
\bibitem [{ope({\natexlab{a}})}]{opennetmap}%
  \BibitemOpen
\bibfield  {journal} {  }\href@noop {} {\enquote {\bibinfo {title} {HIIT
  OpenNetMap project},}\ }\bibinfo {howpublished}
  {[http://opennetmap.rista.fi/]}\BibitemShut {NoStop}%
\bibitem [{ope({\natexlab{b}})}]{opencellid}%
  \BibitemOpen
  \href@noop {} {\enquote {\bibinfo {title} {opencellid},}\ }\bibinfo
  {howpublished} {[http://www.opencellid.org]}\BibitemShut
  {NoStop}%
\bibitem [{loc()}]{locationapi}%
  \BibitemOpen
  \href@noop {} {\enquote {\bibinfo {title} {location-api},}\ }\bibinfo
  {howpublished} {[http://location-api.com]}\BibitemShut {NoStop}%
\bibitem [{\citenamefont {Karikoski}\ and\ \citenamefont
  {Luukkainen}(2011)}]{Karikoski2011c}%
  \BibitemOpen
  \bibfield  {author} {\bibinfo {author} {\bibfnamefont {J.}~\bibnamefont
  {Karikoski}}\ and\ \bibinfo {author} {\bibfnamefont {S.}~\bibnamefont
  {Luukkainen}},\ }in\ \href {\doibase 10.1109/ICIN.2011.6081096} {\emph
  {\bibinfo {booktitle} {Intelligence in Next Generation Networks (ICIN), 2011
  15th International Conference on}}}\ (\bibinfo  {publisher} {IEEE},\ \bibinfo
  {year} {2011})\ pp.\ \bibinfo {pages} {313--318}\BibitemShut {NoStop}%
\bibitem [{\citenamefont {Gan}\ \emph {et~al.}(2007)\citenamefont {Gan},
  \citenamefont {Ma},\ and\ \citenamefont {Wu}}]{Gan2007}%
  \BibitemOpen
  \bibfield  {author} {\bibinfo {author} {\bibfnamefont {G.}~\bibnamefont
  {Gan}}, \bibinfo {author} {\bibfnamefont {C.}~\bibnamefont {Ma}}, \ and\
  \bibinfo {author} {\bibfnamefont {J.}~\bibnamefont {Wu}},\ }\href@noop {}
  {\emph {\bibinfo {title} {Data clustering: Theory, Algorithms, and
  Applications}}},\ \bibinfo {edition} {illustrated edition}\ ed.\ (\bibinfo
  {publisher} {SIAM, Society for Industrial and Applied Mathematics},\ \bibinfo
  {year} {2007})\BibitemShut {NoStop}%
\bibitem [{\citenamefont {Smoot}\ \emph {et~al.}(2011)\citenamefont {Smoot},
  \citenamefont {Ono}, \citenamefont {Ruscheinski}, \citenamefont {Wang},\ and\
  \citenamefont {Ideker}}]{Smoot2011}%
  \BibitemOpen
  \bibfield  {author} {\bibinfo {author} {\bibfnamefont {M.~E.}\ \bibnamefont
  {Smoot}}, \bibinfo {author} {\bibfnamefont {K.}~\bibnamefont {Ono}}, \bibinfo
  {author} {\bibfnamefont {J.}~\bibnamefont {Ruscheinski}}, \bibinfo {author}
  {\bibfnamefont {P.-L.~L.}\ \bibnamefont {Wang}}, \ and\ \bibinfo {author}
  {\bibfnamefont {T.}~\bibnamefont {Ideker}},\ }\href {\doibase
  10.1093/bioinformatics/btq675} {\bibfield  {journal} {\bibinfo  {journal}
  {Bioinformatics (Oxford, England)}\ }\textbf {\bibinfo {volume} {27}},\
  \bibinfo {pages} {431} (\bibinfo {year} {2011})}\BibitemShut {NoStop}%
\bibitem [{\citenamefont {Palla}\ \emph {et~al.}(2005)\citenamefont {Palla},
  \citenamefont {Der\'{e}nyi}, \citenamefont {Farkas},\ and\ \citenamefont
  {Vicsek}}]{Palla2005}%
  \BibitemOpen
  \bibfield  {author} {\bibinfo {author} {\bibfnamefont {G.}~\bibnamefont
  {Palla}}, \bibinfo {author} {\bibfnamefont {I.}~\bibnamefont {Der\'{e}nyi}},
  \bibinfo {author} {\bibfnamefont {I.}~\bibnamefont {Farkas}}, \ and\ \bibinfo
  {author} {\bibfnamefont {T.}~\bibnamefont {Vicsek}},\ }\href {\doibase
  10.1038/nature03607} {\bibfield  {journal} {\bibinfo  {journal} {Nature}\
  }\textbf {\bibinfo {volume} {435}},\ \bibinfo {pages} {814} (\bibinfo {year}
  {2005})}\BibitemShut {NoStop}%
\end{thebibliography}


%

\end{document}